\documentclass[aip, amsmath,amssymb,reprint]{revtex4-2}
\usepackage{graphicx}% Include figure files
\usepackage{dcolumn}% Align table columns on decimal point
\usepackage{bm}% bold math
\usepackage[utf8]{inputenc}
\usepackage[T1]{fontenc}
\usepackage{etoolbox}
\usepackage{xcolor}
\usepackage{amsmath}
\usepackage{amssymb}
\usepackage{float}
\usepackage[caption = false]{subfig}
\usepackage{verbatim}
\usepackage{epstopdf}
\usepackage{bbm}
\usepackage{url}
\usepackage{tikz}
\usepackage{comment}
\usepackage{multirow}
\usepackage{makecell}
\usepackage[pdftex]{hyperref}
\usepackage[normalem]{ulem}
\usepackage[makeroom]{cancel}
\usepackage{soul}

\makeatletter
\def\@email#1#2{%
	\endgroup
	\patchcmd{\titleblock@produce}
	{\frontmatter@RRAPformat}
	{\frontmatter@RRAPformat{\produce@RRAP{*#1\href{mailto:#2}{#2}}}\frontmatter@RRAPformat}
	{}{}
}%
\makeatother
\draft % marks overfull lines with a black rule on the right

\begin{document}

% Use the \preprint command to place your local institutional report number 
% on the title page in preprint mode.
% Multiple \preprint commands are allowed.
%\preprint{}

\title[Generation of directed electron beams]%{Generation of directed electron beams in the MeV range by tight \\ focusing of an ultrashort IR laser in a near-critical plasma}
{Generation of directed electron beams by tight focusing of an ultrashort\\ IR laser in a near-critical plasma}%Title of paper

% repeat the \author .. \affiliation  etc. as needed
% \email, \thanks, \homepage, \altaffiliation all apply to the current author.
% Explanatory text should go in the []'s, 
% actual e-mail address or url should go in the {}'s for \email and \homepage.
% Please use the appropriate macro for the type of information

% \affiliation command applies to all authors since the last \affiliation command. 
% The \affiliation command should follow the other information.

\author{Marianna Lytova}
\email[]{marianna.lytova@inrs.ca}
\affiliation{INRS-EMT, 1650 Boul. Lionel-Boulet, Varennes, Quebec J3X 1P7, Canada}
%\homepage[]{Your web page}
%\thanks{}
%\altaffiliation{}

\author{Fran\c{c}ois Fillion-Gourdeau}
%\email[]{}
\affiliation{INRS-EMT, 1650 Boul. Lionel-Boulet, Varennes, Quebec J3X 1P7, Canada}
\affiliation{Infinite Potential Laboratories, 485 Wes Graham Way, Waterloo, N2L 6R2, ON, Canada}

\author{Simon Valli\`eres}
%\email[]{}
\affiliation{INRS-EMT, 1650 Boul. Lionel-Boulet, Varennes, Quebec J3X 1P7, Canada}

\author{Sylvain Fourmaux}
%\email[]{}
\affiliation{INRS-EMT, 1650 Boul. Lionel-Boulet, Varennes, Quebec J3X 1P7, Canada}

\author{St\'ephane Payeur}
%\email[]{}
\affiliation{INRS-EMT, 1650 Boul. Lionel-Boulet, Varennes, Quebec J3X 1P7, Canada}

\author{Fran\c{c}ois L\'egar\'e}
%\email[]{}
\affiliation{INRS-EMT, 1650 Boul. Lionel-Boulet, Varennes, Quebec J3X 1P7, Canada}

\author{Steve MacLean}
%\email[]{}
\affiliation{INRS-EMT, 1650 Boul. Lionel-Boulet, Varennes, Quebec J3X 1P7, Canada}
\affiliation{Infinite Potential Laboratories, 485 Wes Graham Way, Waterloo, N2L 6R2, ON, Canada}

% Collaboration name, if desired (requires use of superscriptaddress option in \documentclass). 
% \noaffiliation is required (may also be used with the \author command).
%\collaboration{}
%\noaffiliation

\date{\today}

\begin{abstract}
Recent studies have demonstrated the possibility of accelerating electrons to MeV energies in ambient air using tightly focused laser configurations \cite{vallieres2022, Lytova25}. In this article, we explore possible strategies to control and optimize the resulting electron beams using laser and gas parameters. Our theoretical analysis shows that in near-critical plasmas, linearly and circularly polarized pulses are more efficient than radially polarized pulses for electron acceleration. In addition, electron beams obtained from linearly polarized pulses have lower divergence angles. By studying the efficiency of the acceleration process — characterized by the maximum kinetic energy of electrons and their total number — we identify optimal conditions in ambient air at $\lambda_0\approx1.5$ $\mu$m and $a_0\geq15$. We also scale our results to lower-density air and demonstrate that some noble gases (Ne, Ar) are suitable media for accelerating electrons. Our investigations show that this acceleration scheme enables multi-MeV electrons with low divergence using millijoule-class high-repetition rate lasers, making it a promising candidate for applications in medical sciences and ultrafast imaging.

\end{abstract}

% insert suggested PACS numbers in braces on next line

\maketitle %\maketitle must follow title, authors, abstract and \pacs

% Body of paper goes here. Use proper sectioning commands. 
% References should be done using the \cite, \ref, and \label commands
\section{Introduction}
%\label{}
Directed high-energy beams of charged particles are of great interest for medical treatments (see the recent review \cite{Vozenin} and references therein), radiation processing \cite{Cleland:1005393} and many other applications \cite{hamm2012industrial}. In the last few decades, advances in laser physics have offered compelling alternatives to conventional particle accelerators, enabling the development of more compact sources of energetic particles \cite{Daido_2012,RevModPhys.81.1229}. In standard configurations, laser-driven particle sources operate by focusing intense laser pulses onto solid, liquid or gaseous targets. Depending on the composition of the target and the properties of the electromagnetic (EM) beam -- such as its polarization and peak intensity -- various acceleration mechanisms can be triggered,  leading to high energy particles. 
Until now, most of these mechanisms have relied on paraxial beams, as in laser wakefield acceleration (LWFA) \cite{Tajima}, where electrons can reach energies in the range of a few MeV at kHz repetition rates \cite{guenot2017relativistic,PhysRevAccelBeams.21.013401}, and up to a few GeV in the best scenarios, albeit at significantly lower repetition rates\cite{PhysRevSTAB.10.061301,10.1063/1.2718524,PhysRevLett.105.105003,PhysRevX.12.031038,PhysRevLett.133.255001}. Unfortunately, the complex experimental setup required and the difficulty of controlling the beam properties still remain challenges for the implementation of LWFA in medical physics \cite{Nicks,Roa}. Direct laser acceleration (DLA) is another mechanism that requires paraxial configurations and that generates electrons well above the theoretical limit in vacuum by using intricate laser-plasma interactions \cite{PhysRevLett.83.4772,PhysRevLett.79.2053,10.1063/1.873242}.  The optimization of the particle beam properties generated via DLA is still under development \cite{Hussein_2021,doi:10.1126/sciadv.adk1947}.

Meanwhile, significant theoretical advances have predicted efficient electron acceleration driven by \textit{non-paraxial} laser pulses\cite{PhysRevLett.88.095005,PhysRevE.68.046407,10.1063/1.2830651,varin2002acceleration,Bochkarev_2007,10.1063/1.2830651,app3010070,PhysRevLett.111.224801}. Recent experiments have confirmed the potential of such non-paraxial configurations for relativistic electron acceleration, particularly when employing tightly focused field geometries in both near-critical density plasmas \cite{vallieres2022} and highly underdense plasmas \cite{Payeur,PhysRevAccelBeams.19.021303,powell2024relativistic}.
In these experiments, the detected maximum kinetic energy of electrons reaches up to 1.4 MeV using mJ-class lasers and the generated beams exhibit highly directional propagation. 

\begin{figure}[h]
	\centering{
		\includegraphics[scale=0.8,trim={20 40 20 40},clip]{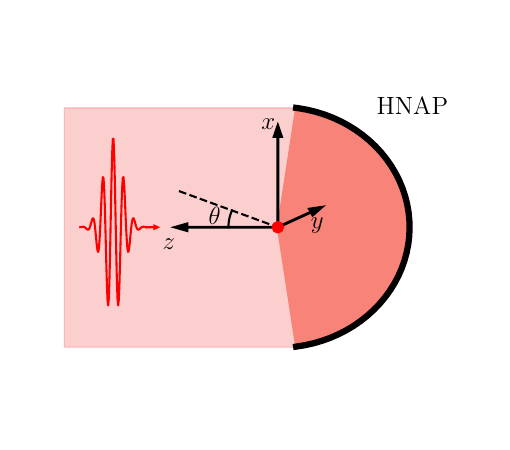}}
	\caption{Geometry of the HNAP reflector: tight focusing of the incident ultrashort IR laser pulse. Here $z$ is the direction of propagation of the reflected focused pulse, $(x, y)$ is the polarization plane.}  \label{hnap}
\end{figure}

In a previous article\cite{Lytova25}, we presented the results of 3D particle-in-cells (PIC) simulations for the setup of the experiment using a High Numerical Aperture Parabola (HNAP)\cite{Vallieres:23} focused in ambient air\cite{vallieres2022}. In this experimental setup, displayed in Fig. \ref{hnap},  the incoming infrared (IR) laser pulse was focused tightly by an on-axis high numerical aperture parabolic mirror. In a small region near the focus, the field strength becomes sufficiently high to ionize nitrogen and oxygen atoms, thus releasing electrons that are subsequently accelerated.
We demonstrated that in this configuration, the electron acceleration is due to the direct action of the relativistic ponderomotive force (PMF), despite the fact that a near-critical plasma is created. Here, the term ``direct'' is used to emphasize the role of the PMF and to distinguish this mechanism from LWFA, which also requires the PMF but in a different capacity.
In this work, based on the model presented in our previous article\cite{Lytova25}, we explore direct ponderomotive acceleration of electrons by ultrashort tightly focused pulses in near-critical plasma, taking into account different laser setups and gas compositions. Our goal is to optimize these conditions to improve the electron beam properties, namely, to increase the maximum energy and charge while maintaining directionality.

This article is separated as follows. In Section \ref{sec:pmf}, the PMF is reviewed and applied qualitatively to vacuum electron acceleration in tightly focused fields. In Section \ref{plasma}, the theoretical model based on the PIC method is presented and numerical results are obtained for different laser and gas parameters. Different target gases are investigated in Section \ref{sec:gases}. We conclude with additional perspectives in section \ref{sec:conclusion}.

\section{Acceleration of electrons by relativistic PMF in vacuum}
\label{sec:pmf}

In this section, some general features of the relativistic PMF are reviewed and analyzed. Then, some results of this analysis are applied to vacuum electron acceleration. The main goal here is to highlight potential differences between vacuum acceleration and plasma effects introduced in the next sections. 

\subsection{Relativistic PMF}
In the relativistic regime, the PMF depends not only on the inhomogeneity of the electric field, but also on the magnetic part of the Lorentz force and its inhomogeneity. Complex laser field configurations with space-time couplings, such as tightly-focused beams, complicate or may even preclude the derivation of analytical formulas for PMF. To observe the parameter dependencies of these beams, we have to rely on solutions obtained for ideal conditions, while recognizing the simplification of this approach. 

\begin{figure}[h]
	\centering{
		\includegraphics[scale=1,trim={15 30 40 25},clip]{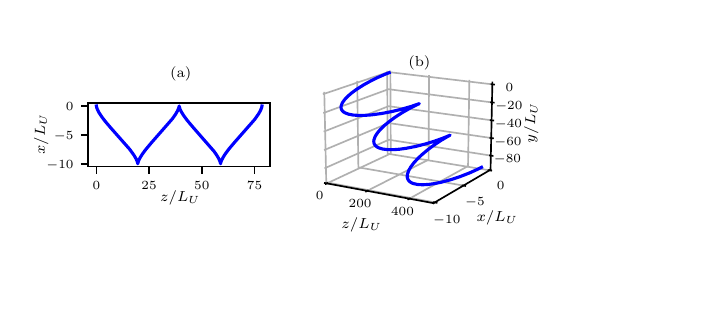}}
	\caption{Trajectories of an electron in a vacuum in the field of a monochromatic plane (a) LP or (b) CP wave. Here the vector potential $a_x=-a_0\sin(\omega t-kz)H(\omega t-kz)$ and also for (b) $a_y=a_0\cos(\omega t-kz)H(\omega t-kz)$, $H$ stands for the Heaviside step function, $a_0=5$, coordinate are normalized by length $L_U=\lambda/2\pi$. Initial conditions: $x(0)=y(0)=z(0)=0$, $p_x(0)=p_y(0)=p_z(0)=0$.}  \label{traj}
\end{figure}

As a first approximation, we turn to the classical analytical solution for the motion of a charged relativistic particle in a plane monochromatic EM wave polarized in the $x$-direction \cite{Landau}. When such a linearly polarized (LP) wave  of frequency $\omega$ and wavelength $\lambda$ with vector potential $A_x(\omega t-kz)$ (here $k=2\pi/\lambda$) propagates in the $z$-direction, the electron momentum has not only a component $p_x=a_x$ oscillating with the electric field, but also acquires a component in the direction of wave propagation $p_z=a_x^2/2 \geq 0$ \cite{Stupakov}. Here and in the following, the momentum is expressed in units of $m_ec$, and $a_x$ is the vector potential, normalized to $e/m_ec$ with amplitude $a_0=eE_0/m_ec\omega$, where $E_0$ is the  maximum field strength.  Similarly, the general solution for circularly polarized (CP) waves include a forward acceleration in the direction of propagation $p_z=a_0^2-a_0a_y \geq 0$, on top of 
oscillations with transverse components $p_x=a_x$ and $p_y=a_y-a_0$. Figure~\ref{traj} shows the trajectory of an electron accelerated in (a) LP or (b) CP monochromatic wave propagating in vacuum. In both cases, the most significant regular motion occurs in the $z$-direction.

For cases far from a plane wave with strong inhomogeneities, such as the ones considered latter in this article, the analytical description is based on the cycle-averaged solution in the oscillation center system \cite{Landau} and the Lagrange equation of motion for this oscillating center.
Then the relativistic PMF acting on an electron with oscillation center speed ${\bf v}_0$ in a monochromatic traveling wave ${\bf A}({\bf r},t)=\text{Re}\{\hat{\bf A}({\bf r})e^{i\omega t}\}$ is \cite{Bauer}
\begin{align}\label{RPMF}
	{\bf F}_p = -\dfrac{m_ec^2}{\gamma_0} \Big\{\nabla^C\gamma_{a}+\dfrac{\gamma_0-1}{v_0^2}\big({\bf v}_0\cdot\nabla^C\gamma_a \big){\bf v}_0\Big\},
\end{align}
where $\gamma_a = \sqrt{1+e^2|\hat{\bf A}({\bf r})|^2/\alpha m_e^2c^2} = \sqrt{1+|{\bf a}({\bf r})|^2/\alpha}$ with parameter $\alpha=2$ for the LP case and $\alpha=1$ for CP. Also $\gamma_0 = 1/\sqrt{1-v_0^2/c^2}$ and the index $C$ stands for the comoving system. Based on \eqref{RPMF}, we can obtain an expression for the $z$-component of the PMF  when ${\bf v}_0$ is aligned with the $z$-axis, the direction of the predominant drift of the particles, as we saw from the analytical solutions in vacuum presented  Fig.~\ref{traj} and previous simulations \cite{Lytova25}, then
\begin{align}\label{PMF_z}
	F_{pz}\approx-m_ec^2\dfrac{d}{dz}\sqrt{1+\dfrac{|{\bf a}({\bf r})|^2}{\alpha}}.
\end{align}
Thus,  when $|{\bf a}({\bf r})|$ changes from $a_0$ to 0 as a function of the position ${\bf r}$, we can write the kinetic energy averaged over one cycle as
\begin{align}\label{ekin}
	\langle\mathcal{E}_K\rangle=m_ec^2\left(\sqrt{1+\frac{ a_0^2}{\alpha}}-1\right).
\end{align}

Although formulas \eqref{PMF_z} and \eqref{ekin} have a clear physical meaning and can be useful for a qualitative understanding of some acceleration processes, we cannot completely rely on them when studying the  electron acceleration in ultrashort tightly focused laser beams in plasmas. First, we cannot separate the temporal and spatial parts for this type of pulses\cite{April_2010, Jolly}, so the analytical procedures leading to the above solutions are not applicable. Next, these formulas should only work when  electron oscillates with the field for many periods, while we consider here few cycle pulses. Finally, interaction with plasma can strongly change the pulse shape and thus, affect the acceleration process. Nevertheless, we compare our  simulation with these formulas in Sec.~\ref{plasma} to benchmark the numerical results and to understand the dominant acceleration mechanisms at play.

\subsection{Tightly focused laser beams}\label{sec:laser_model}

The theoretical approach described in our previous study\cite{Lytova25}, which we use further in Sec.~\ref{plasma} for simulations in plasma, requires  a model for a tightly focused electromagnetic field in vacuum. The latter is introduced in PIC simulations as an initial condition, outside the gas target and before the laser-matter interaction begins. It can be chosen as a numerical solution to take into account the detailed features of the focusing system  or an analytical solution with suitable parameters, such as the ones obtained using the April model\cite{April_2010}. Originally developed for radially polarized (RP) pulses~\cite{April_2010}, this model was extended to linearly polarized (LP) pulses in our previous work~\cite{Lytova25}, and we further adapt it here to circularly polarized (CP) pulses in Appendix~\ref{exact_CP}. These analytical solutions are in satisfactory agreement with a more accurate numerical method based on Stratton-Chu integrals \cite{Dumont_2017}.

In the experiment considered\cite{vallieres2022}, tight focusing of the pulse occurs using a HNAP reflector, see Fig.~\ref{hnap}. Thus, we adopt a numerical aperture of NA=0.95 in our simulations, see Eq.~\eqref{b_NA}. 
For all polarization types, we consider ultrashort pulses with time duration $\tau_\text{FWHM}$, see Eq.~\eqref{tau_w0}, such that  the number of cycles in a pulse is the same for any central wavelength $\lambda_0$: $\tau_\text{FWHM}\cdot\omega_0=\mathrm{const}=4\pi$, where $\omega_0=2\pi/\lambda_0$. 
\begin{figure}[h]
	\centering{
		\includegraphics[scale=0.88,trim={10 0 0 12},clip]{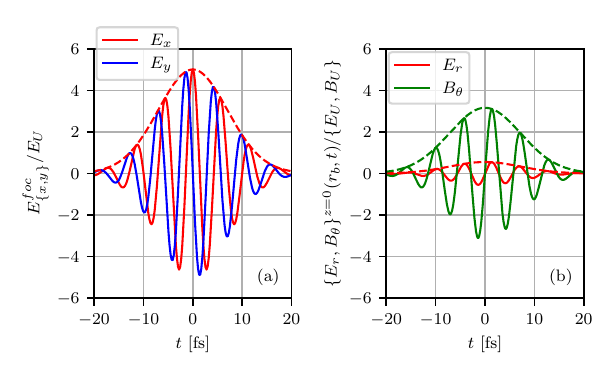}}
	\caption{Analytical time solutions for the tightly focused laser beam models at the focal plane $z=0$. For (a) linear ($E_x$) and circular ($E_x$, $E_y$) polarization, the field is measured at the focus  $x=y=0$, while for (b) radial polarization, it is measured at the position $r=3.25b$, where $b$ is the confocal parameter. The other parameters are set to $\lambda_0$ = 1.8 $\mu$m, NA=0.95 ($b=0.2$ $\mu$m), $s$ = 57.18,  $a_0$ = 5, see Eqs.~\eqref{b_NA}, \eqref{tau_w0} and text there.}  \label{env_cp_rp}
\end{figure}
Some examples of LP and CP pulses are displayed in Fig.~\ref{env_cp_rp}~(a), where the maximum amplitude $a_0=E_{x0}/E_U=5$ is achieved at the focal point $x=y=z=0$. We normalized the electromagnetic field as $\boldsymbol{B}/B_U$ and $\boldsymbol{E}/E_U$, where the unit electric and magnetic fields are $E_U=m_ec\omega_0/e$ and $B_U=m_e\omega_0/e$, respectively. 

\subsection{Electrons in tightly focused fields}\label{predictions}

From the field solutions shown in Fig.~\ref{env_cp_rp}~(a), one can expect direct electron acceleration due to the PMF: $F_{pz}^{\mathrm{LP}}= v_xB_y$ or $F_{pz}^{\mathrm{CP}}= v_xB_y-v_yB_x$, where $B_x/B_U=-E_y/E_U$ and $B_y/B_U=E_x/E_U$ as can be seen from Eqs.~\eqref{Ex}-\eqref{Bz}. As we know, these forces induce an average motion going in the $z$-direction, see Fig.~\ref{traj}.

The situation looks different for RP pulses. First of all, now the only non-zero component on the axis $z$ (at $r=0$) is $E_z$, which has an amplitude $a_0=5$. This field component can also accelerate particles, but in very rarefied plasma  \cite{Payeur, powell2024relativistic}. For ponderomotive acceleration in the $z$ direction, a combined action of the radial $E_r$ and azimuthal $B_\theta$ fields is required, which in the focal plane $z=0$ increases from 0 at $r=0$ to a maximum value at $r\approx3.25b$ ($b$ is the confocal parameter \eqref{b_NA}), and then asymptotically decreases to 0 with  increasing $r$. As can be estimated from looking at amplitudes of $E_r$ and $B_\theta$ in Fig.~\ref{env_cp_rp} (b), the maximum PMF in the $z$-direction for RP is more than an order of magnitude lower than for LP or CP pulses. 
Moving away from the $z$-axis by a distance $b$, we can find slightly larger amplitudes for $E_r$ (up to $\approx1$), but there its oscillations are already noticeably desynchronized with the $B_\theta$ oscillations, so we cannot expect much benefit from such an increase in the amplitude of the electric field.

\begin{figure}[h]
	\centering{
		\includegraphics[scale=0.89,trim={12 0 0 0},clip]{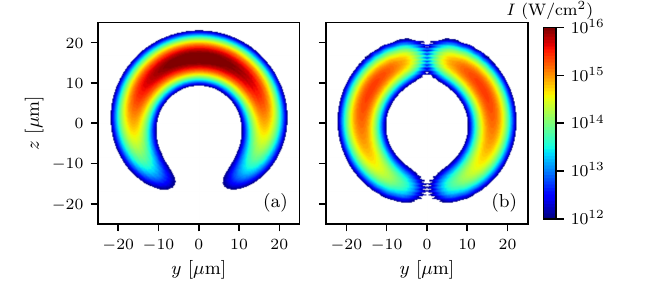}}
	\caption{Propagation of the intensity envelope in vacuum for the cases of (a) CP and (b) RP tightly-focused ultrashort pulses. Direction of propagation is $z$, time $t=27.5$ fs after focusing, laser strength parameter $a_0=5$, central wavelength $\lambda_0=1.8$ $\mu$m, envelope duration $\tau_\text{FWHM}=12$ fs.}  \label{vac_cp_rp}
\end{figure}

Now let us consider the contribution to the PMF with strong field inhomogeneities and see how it depends on polarization. Figure~\ref{vac_cp_rp} shows analytical solutions propagating in the $z$-direction: (a) - CP pulse and (b) - RP pulse.
For CP pulses, see Fig.~\ref{vac_cp_rp} (a), as well as for LP pulses \cite{Lytova25}, the PMF is directed against and proportional to the value of the intensity gradient $-\nabla|{\bf E}|^2$: forward at the leading edge and backward at the trailing edge, which facilitates the directionality of the electron beam. In contrast, the field geometry shown in Fig.~\ref{vac_cp_rp} (b) results in a defocusing of the accelerated electrons due to the PMF deflecting them away from the $z$-axis. 

Thus, we see clear advantages of LP and CP tightly focused pulses compared to RP ones for accelerating directed electron beams due to ponderomotive force. Can we now expect the CP pulses to be more efficient as an accelerator than the LP pulses? If we apply the formula \eqref{ekin}, we come to the conclusion that the cycle-averaged energy for CP pulses should be significantly higher, e.g. for $a_0=5$ it is 2.1 MeV vs 1.4 MeV in the LP case. However, as can be seen from Fig.~\ref{env_cp_rp}, there are only a few cycles when the electrons oscillate with the field, so the conditions for \eqref{ekin} to be valid are not satisfied, so we expect some discrepancies. We return to this question in Sec.~\ref{polar}, where we compare the results of numerical experiments. 

To conclude this section, we recall one more fact related to the polarization of the ionizing field. The electron drift velocity during ionization contributes to the spectrum, and in the case of the CP laser field this contribution is stronger due to the drift perpendicular to the component of the ionizing field \cite{Corkum}. For example, we can see this drift along the $y$-axis in Fig.~\ref{traj} (b) for the given initial conditions. One could expect that in a plasma, where many atoms in randomly distributed states are ionized in different phases of the wave, these drifts contribute to all perpendicular components of the momentum.
In the next section, we will numerically simulate the acceleration of electrons in tightly focused fields of different polarization to see how this and other factors affect the spectra.

\section{Simulations of electron acceleration in air plasma}\label{plasma}
The model we developed for simulating electron acceleration in dense plasma is described in detail in a previous article~\cite{Lytova25}, here we briefly list the key points. Our simulations are performed using the Smilei PIC code \cite{Smilei18}, in 3D Cartesian geometry with the laser beam propagation direction along the $z$-axis.
Due to the spherical shape of the tightly focused field phase front, see e.g. Fig.~\ref{vac_cp_rp}, we use a spherical gas target with a smooth boundary filled with initially neutral atoms:
\begin{align}
    n_{gas}(x,y,z)=n_{gas0}\exp\Big\{-\Big(\dfrac{x^2+y^2+z^2}{2\sigma^2}\Big)^p\Big\},
\end{align}
where super-Gaussian order $p=6$ and in the case of air under standard atmospheric conditions $n_{air0}=2.687\times10^{19}$~cm$^{-3}$.
The size of the target $\sigma=30[\mu\text{m}]~\cdot~\lambda_0[\mu\text{m}]/1.8\mu\text{m}$ is chosen so that the ionization at the boundary is small and the field energy there is insufficient for inducing noticeable particle acceleration. We set the analytical solution for tightly focused field with a certain polarization around the gas target as an initial condition and let it propagate in the plasma according to the fourth order accurate Maxwell's equation solver of the PIC code on a grid $\Delta x=\Delta y = \Delta z=\lambda_0/32$ and with a time step $\Delta t=0.99\Delta z/\sqrt{3}c$. As this pulse moves through the target, it ionizes the air (Smilei PIC code includes a semi-classical ADK ionization model \cite{Nuter}) and accelerates the resulting electrons. 

For diagnostics, we measure the energy-angular distribution of accelerated electrons that cross some specified frontiers $\pm10\cdot\lambda_0$ -- before or after the focus -- at several pre-selected time points, including the time of focusing in vacuum. At the same time points, we store the spatial distributions of the number density and kinetic energy density of electrons, the charge states of ions, and the EM field. Also, some scalar characteristics, such as kinetic and electromagnetic energy integrated over the simulation block, are also available for analysis in the Smilei PIC code and are recorded.

Finally, we note that our simulations were conducted for various wavelengths corresponding to known high energy laser technologies. For example, 800 nm corresponds to a Ti:Sapphire laser, while 1.47 $\mu$m and 1.8 $\mu$m can be achieved with Optical Parametric Amplifier pumped by an Ytterbium technology or a Ti:Sapphire system \cite{Thire2015_10mJ5cycle}, respectively. The short-wave infrared range up to 2.5 $\mu$m is achievable with a Tm-doped fiber laser system \cite{Heuermann:22}, while wavelengths around 10 $\mu$m can be obtained with CO$_2$ lasers.

\subsection{Dependence of electron spectra on laser pulse polarization}\label{polar}
\begin{figure}[h]
	\centering{
		\includegraphics[scale=1,trim={0 0 0 0},clip]{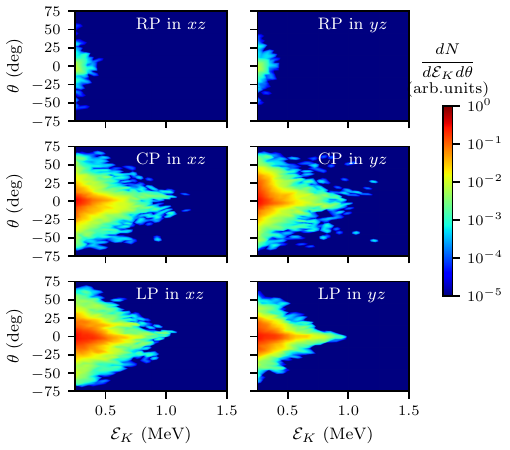}}
	\caption{Energy-angular spectra of electrons accelerated forward by a tightly focused RP or CP or LP (last in direction $x$) pulse with $\lambda_0=1.8$ $\mu$m and $a_0 = 5$ are shown at time $t = 91.5\cdot\lambda_0[\mu\text{m}]$ fs in $xz$ and $yz$ planes.}  \label{spec_rpcplp}
\end{figure}

We start by looking at the effect of the polarization on electron spectra. This is displayed in  Fig.~\ref{spec_rpcplp}, where the forward spectra of electrons accelerated in tightly focused RP, CP and LP pulses are shown. The central wavelength is $\lambda_0=1.8$ $\mu$m while the field strength is $a_0=5$, corresponding to experimental conditions for ambient air acceleration using LP pulses\cite{vallieres2022}. As we discussed in section~\ref{predictions} for relativistic PMF acceleration, the electron energy gain for RP pulses is expected to be less than that for LP and CP pulses. The same trend is observed in the full simulations that include the plasma, as confirmed by the results of Fig.~\ref{spec_rpcplp}. The maximum kinetic energy reaches $\approx 0.35$ MeV for RP pulses while it goes up to $\approx 1$ MeV for CP and LP pulses. The nature of the acceleration mechanism, i.e. direct acceleration of the PMF, is also confirmed by the predominantly forward acceleration, with a significantly smaller number and maximum energy of backward electrons – in all three cases less than 0.35 MeV. Interestingly, similar observations were made experimentally for another experimental setup when studying vacuum acceleration of electrons using plasma mirrors, where LP pulses produced electron energies three to four times higher than RP pulses \cite{PhysRevX.10.041064}.

\begin{figure}[h]
	\centering{
		\includegraphics[scale=1,trim={0 0 0 0},clip]{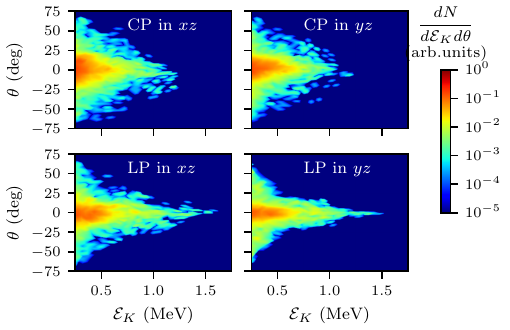}}
	\caption{Energy-angular spectra of electrons accelerated forward by a tightly focused CP or LP (last in direction $x$) pulse with $\lambda_0=2.5$ $\mu$m and $a_0 = 5$ are shown at time $t = 91.5\cdot\lambda_0[\mu\text{m}]$ fs in $xz$ and $yz$ planes.}  \label{spec_cplp}
\end{figure}

\begin{figure}[t]
	\centering{
		\includegraphics[scale=1,trim={10 0 0 0},clip]{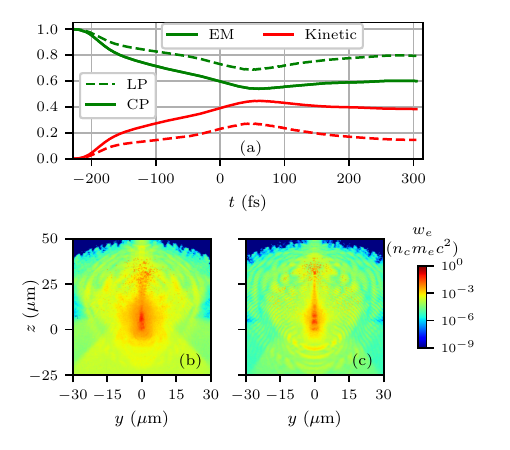}}
	\caption{(a) Normalized EM energy and kinetic energy of electrons over time for acceleration by a laser pulse with central wavelength $\lambda_0 = 2.5$ $\mu$m, strength $a_0=5$ and LP or CP polarization.
	Bottom row: distributions of electrons kinetic energy $w_e(y,z)$ in the field of a (b) CP and (c) LP tightly-focused laser pulse at time $t = 61.2\cdot\lambda_0[\mu\text{m}]$ fs.}  \label{KinDense25}
\end{figure}

\begin{figure}[h]
	\centering{
		\includegraphics[scale=1,trim={0 0 0 0},clip]{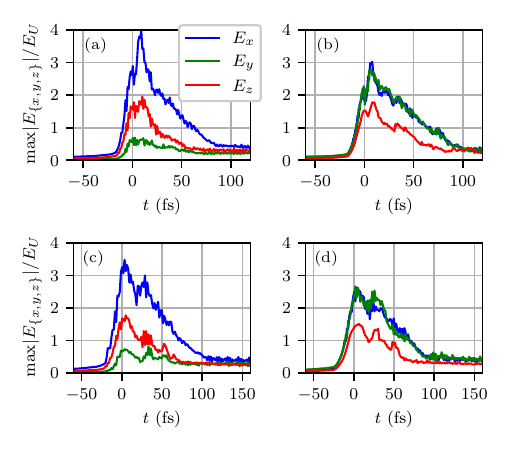}}
	\caption{Electric field maxima in the simulation box over time for (a), (c) LP and (b), (d) CP pulses with $a_0$ = 5 and wavelength (a), (b) $\lambda_0=1.8$ $\mu$m and (c), (d) $\lambda_0=2.5$ $\mu$m.}  \label{absMax25}
\end{figure}

However, the prediction based on Eq.~\eqref{ekin} that electron kinetic energy should be higher for acceleration by CP pulses compared to LP pulses is not fully confirmed. The energy maxima are similar in both cases, although the CP pulse additionally produces some randomly distributed ``flakes'' of higher-energy particles. Moreover, as seen in Fig~\ref{spec_cplp}, for $\lambda_0=2.5$ $\mu$m, the maximum kinetic energy of electrons is clearly higher for LP pulses.
The discrepancy between these simulation results and Eq.~\eqref{ekin} can be attributed to the fact that the electrons are not accelerated by a free field in vacuum, but rather by a self-consistent field in a plasma with a density close to the critical value
\begin{align}
\label{eq:nc}
   n_c=4\pi^2\varepsilon_0m_ec^2/e^2\lambda_0^2 .
\end{align}
At such densities, the ionized electrons can effectively acquire kinetic energy by interacting with EM waves, see Fig.~\ref{KinDense25}. 
Since the electric field is always ``on'' for the CP pulse, the ionization during its propagation to the focus is stronger than for field ionization by the LP pulse. However, this does not provide an advantage in accelerating the directed electron beams, since by the time the focal position is reached, both components $E_x$, $E_y$ of the CP pulse are attenuated more than the component $E_x$ of the LP pulse, see Fig.~\ref{absMax25}.
Although the conversion of energy into electron motion is stronger for CP pulses, this energy is mainly spent on heating the plasma around the pulse, see Fig.~\ref{KinDense25}(b),(c), whereas the achieved energy maxima and the number of the most energetic electrons in the spectrum tail show the advantage of the LP pulse.  
\begin{figure}[t]
	\centering{
		\includegraphics[scale=1.1,trim={5 15 20 25},clip]{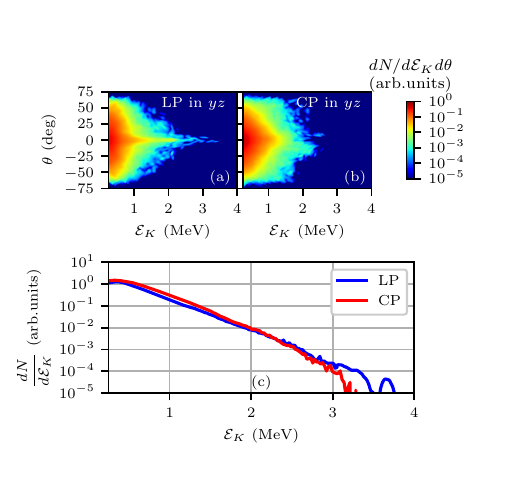}}
	\caption{Upper row: Energy-angular spectra of electrons accelerated forward by a tightly focused (a) LP (in direction $x$) or (b) CP laser pulse with wavelength $\lambda_0=1.47$ $\mu$m and strength $a_0 = 12$ are shown at time $t = 76.2\cdot\lambda_0[\mu\text{m}]$ fs in $yz$ plane; (c) Angle-averaged energy spectra for the same parameters as for (a) and (b).}  \label{a0_12}
\end{figure}
Note that with increasing $a_0$ this trend persists, it even becomes more noticeable as seen on Fig.~\ref{a0_12}: the angular spectrum is wider in the case of CP pulses, but the maximum energy is higher when accelerated by an LP pulse, as is the number of accelerated particles with an energy of more than 3 MeV. Thus, our simulation results show that for the same $a_0$ and $\lambda_0$, the LP pulse is more effective both in imparting a higher maximum kinetic energy to the electron beam and in maintaining the beam directionality. For these reasons, in further simulations we consider acceleration only in LP pulses.

\subsection{Control of electron beam generation in air plasma}\label{ambient_air}
In this section we study the influence of the laser parameters $\lambda_0$ and $a_0$ on the resulting electron spectra in ambient air. At first, let us estimate the cycle-averaged kinetic energy using formula \eqref{ekin} for LP pulse with $a_{0x}\approx4$, $a_{0y}\approx0.7$ (finite-time pulse cannot be purely LP) and for CP pulse with $a_{0x}\approx3$, $a_{0y}\approx2.8$, which corresponds to data from Fig.~\ref{absMax25}(a),(b). From here we obtain fairly close values: $\langle\mathcal{E}_K^\text{LP}\rangle \approx 1.04$ MeV and $\langle\mathcal{E}_K^\text{CP}\rangle \approx 1.06$ MeV, which are also consistent with Fig.~\ref{spec_rpcplp}. Such simple estimates do not work for the fields shown in Fig.~\ref{absMax25}(c),(d), since they lead to $\langle\mathcal{E}_K^\text{LP}\rangle \approx 0.88$ MeV and $\langle\mathcal{E}_K^\text{CP}\rangle \approx 0.93$ MeV which are far below the simulated results for $\lambda_0=2.5$ $\mu$m and $a_0=5$, with values of 1.6 MeV and 1.28 MeV, respectively, see Fig.~\ref{spec_cplp}. So, we are faced with the (albeit expected) inability of the equation~\eqref{ekin} to describe the acceleration of electrons by an ultrashort and tightly focused pulse, even if we take into account the attenuation of the wave in the plasma.

\begin{figure}[h]
	\centering{
		\includegraphics[scale=1.1,trim={5 15 20 25},clip]{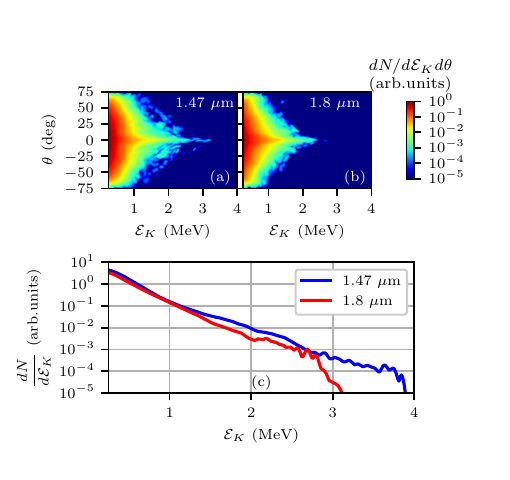}}
	\caption{Upper row: Energy-angular spectra of electrons accelerated forward by a tightly focused LP (in direction $x$) laser pulse with strength $a_0 = 15$ and wavelength (a) $\lambda_0=1.47$ $\mu$m or (b)  $\lambda_0=1.8$ $\mu$m are shown at time $t = 91.5\cdot\lambda_0[\mu\text{m}]$ fs in $yz$ plane; (c) Angle-averaged energy spectra for the same parameters as for (a) and (b).}  \label{spec_a0_15}
\end{figure}

To explain this, let us now recall one more parameter that needs to be taken into account when setting up a simulation or experiment. In ambient air plasmas with local average ionization states of nitrogen $Z_\mathrm{N}$ and oxygen $Z_\text{O}$, the local electron number density can be written approximately as~\cite{Lytova25} 
\begin{align}
	n_e({\bf r})/n_c= n_{e1}\big[0.79Z_\text{N}({\bf r}) + 0.21Z_\text{O}({\bf r})\big],
\end{align}
where under standard conditions, assuming dissociation of N$_2$ and O$_2$,  $n_{e1}\approx0.048[\lambda_0(\mu\text{m})]^2$ is a coefficient that has the meaning of the ratio of the electron density to the critical density in the case of single ionization: $Z_\text{N}=Z_\text{O}=1$. The group velocity of the EM pulse in the plasma is $v_\text{gr}=c\sqrt{1-\omega_{pe}/\omega^2}=c\sqrt{1-n_e/n_c}$, therefore the pulse cannot enter deeply into the region with such number density, but is scattered at the boundary. The situation is even more complicated because we are dealing with multiple regions of critical plasma, not just one in focus. However, we can calculate the average value of the ratio $n_e/n_c$ in the vicinity of the focus, for the volume $5\lambda_0\times5\lambda_0\times5\lambda_0$. This data is presented in Table~\ref{tab1} for some spectra already shown in Figs.\ref{spec_rpcplp},\ref{spec_cplp} and \ref{a0_12}, while the last 2 columns of the table are extracted from spectra in Fig.~\ref{spec_a0_15}.
\begin{table}[h]
	 \caption{\label{tab1} Electron number density averaged over a near-focal region of size $5\lambda_0\times5\lambda_0\times5\lambda_0$ in critical units and m$^{-3}$ for ionization and acceleration by a LP laser pulse with different combinations of laser parameters $a_0$ and $\lambda_0$ and corresponding pulse duration $\tau_\text{FWHM}$ and energy per pulse $\mathcal{E}_L$, see Appendix~\ref{exact_CP}. The last row shows the maximum values of electron kinetic energies found in the simulations.}
%\begin{ruledtabular}
	\begin{tabular}{cc|cc|c|cc}
        \hline \hline
		\multicolumn{2}{c|}{$a_0$}                     & \multicolumn{2}{c|}{5}    & 12 & \multicolumn{2}{c}{15}    \\ \hline
		\multicolumn{2}{c|}{$\lambda_0$ ($\mu$m)}                     & \multicolumn{1}{c|}{1.8} & 2.5 &1.47  & \multicolumn{1}{c|}{1.47} &1.8  \\ \hline
        \multicolumn{2}{c|}{$\tau_\text{FWHM}$ (fs)}                     & \multicolumn{1}{c|}{12} & 16.7 &9.8  & \multicolumn{1}{c|}{9.8} &12  \\ \hline
        \multicolumn{2}{c|}{$\mathcal{E}_L$ (mJ)}                     & \multicolumn{1}{c|}{1} & 1.4 &4.7  & \multicolumn{1}{c|}{7.3} &9  \\ \hline
		\multicolumn{1}{c|}{\multirow{2}{*}{$n_e/n_c$}} & $t=0$ & \multicolumn{1}{c|}{0.35} & 0.56 & 0.41 & \multicolumn{1}{c|}{0.44} & 0.63 \\ \cline{2-7} 
		\multicolumn{1}{c|}{}                  & $t\geq57/\omega_0$ & \multicolumn{1}{c|}{0.48} &0.8  & 0.51 & \multicolumn{1}{c|}{0.52} &0.77  \\ \hline
		\multicolumn{1}{c|}{\multirow{2}{*}{	\makecell{$n_e$ \\ $(\times10^{20}$ cm$^{-3}$)} }} &$t=0$  & \multicolumn{1}{c|}{1.21} &1.0  & 2.14 & \multicolumn{1}{c|}{2.28} & 2.18 \\ \cline{2-7} 
		\multicolumn{1}{c|}{}                  & $t\geq57/\omega_0$ & \multicolumn{1}{c|}{1.66} & 1.43 & 2.62 & \multicolumn{1}{c|}{2.70} & 2.64 \\ \hline
		\multicolumn{2}{c|}{$\mathcal{E}_K^\text{max}$ (MeV)}                     & \multicolumn{1}{c|}{1.05} & 1.6 &3.75  & \multicolumn{1}{c|}{3.15} &3.9  \\
        \hline \hline
	\end{tabular}
%\end{ruledtabular}
\end{table}

These results indicate that the underlying dynamics may be more intricate than they first appear. For example, as is known from previous simulations \cite{Lytova25}, the maximum kinetic energy of the electron beam is achieved when accelerated by a pulse with $\lambda_0=2.5$ $\mu$m for a laser tuned to $a_0=5$ (meaning the maximum in vacuum). In this case, the ratio $n_e/n_c$ near the focal region increases to 0.8, see Table~\ref{tab1}. However, if we consider the cases with $a_0=15$ and two wavelengths $\lambda_0=1.47$ $\mu$m and $\lambda_0=1.8$ $\mu$m, we see the opposite result: although the $n_e/n_c$ ratio is still higher for the larger $\lambda_0$: 0.77 versus 0.52, the maximum energy is higher for the smaller $\lambda_0$, see Fig.~\ref{spec_a0_15}. If we take into account the absolute electron density $n_e$, then in both comparisons: 1.8 $\mu$m versus 2.5 $\mu$m and 1.47 $\mu$m versus 1.8 $\mu$m, it is higher for the lower $\lambda_0$, since then we have higher peak intensity and hence ionization. Thus, we cannot ascribe the lower number of electrons (implying worse statistics)  as the reason for the lower maximum kinetic energy - for 2.5 $\mu$m this parameter is also lower than for 1.8 $\mu$m, although the maximum kinetic energy is still higher.
So, we see that the average $n_e/n_c$ ratio near the focus is not an accurate predictor of the efficiency of the acceleration process. From the simulations, we can only conclude that to achieve higher kinetic energies and at the same time a noticeable number of accelerated electrons, the average $n_e/n_c$ parameter should be in the range of $0.5-0.8$ for the laser parameters considered. At lower values, the total number of accelerated particles is significantly smaller, although some of them can still reach quite high energies. Conversely, in a medium where $n_e/n_c \geq 1$ the pulse decays too quickly to accelerate electrons to appreciable energies. 

\begin{figure}[h]
 	\centering{
 		\includegraphics[scale=1,trim={0 0 0 0},clip]{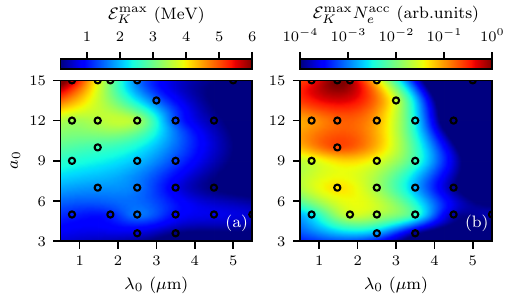}}
 	\caption{(a) The maximum kinetic energy of accelerated electrons $\mathcal{E}_K^\mathrm{max}$ and (b) the product of $\mathcal{E}_K^\mathrm{max}$ and the total number of electrons with kinetic energy $\geq$ 250 keV $N_e^\mathrm{acc}$ are shown on the plane $(\lambda_0, a_0)$ assuming that the medium is air at standard conditions.The circles represent the results of a 3D PIC simulation to which RBF interpolation with a thin plate spline kernel was applied.}  \label{a0_l0_plane}
\end{figure}

To summarize our results, Fig.~\ref{a0_l0_plane} shows the maximum kinetic energy $\mathcal{E}_K^\mathrm{max}$ and the ``acceleration efficiency'' defined as $\mathcal{E}_K^\mathrm{max}N_e^\mathrm{acc}$ in the plane $(\lambda_0, a_0)$. Here $N_e^\mathrm{acc}$ is the total number of accelerated electrons with kinetic energies greater than 250 keV. 
To obtain these plots, we used radial basis function (RBF) interpolation over more than 20 points obtained from 3D PIC simulation results. Not all of these points were calculated at the high spatial resolution that we typically use to obtain detailed energy-angle spectra \cite{Lytova25}, see Sec.~\ref{plasma}, but even doubling the step size was still sufficient to obtain reasonably accurate integrated spectra while allowing significant reductions in computational resources. From Fig.~\ref{a0_l0_plane}(a), (b) it can be seen that the best parameters for maximizing the kinetic energy of electrons are shifted to lower $\lambda_0$ relative to the best parameters for maximizing the acceleration efficiency. In particular, a tightly focused laser configuration with $a_0=15$ and $\lambda_0=0.8$~$\mu$m can accelerate electrons to $\mathcal{E}_K^\mathrm{max}=5.7$~MeV, whereas with the same $a_0$ but $\lambda_0=1.47$~$\mu$m, we get $\mathcal{E}_K^\mathrm{max}=3.9$ MeV. However, the number of accelerated electrons in the first case is 5 times less than in the second.
This can be explained in terms of the $n_e/n_c$ ratio. Indeed $n_{e1}(0.8 \mu\mathrm{m})\approx0.03$, while $n_{e1}(1.47 \mu\mathrm{m})\approx0.11$. When ionizing with $a_0=15$ at different wavelengths, it leads to $Z_\mathrm{N}\approx5$ and $Z_\mathrm{O}\approx6$ - for $\lambda_0=0.8$ $\mu$m and $Z_\mathrm{N}\approx4.9$ and $Z_\mathrm{O}\approx5.5$ - for $\lambda_0=1.47$ $\mu$m. Finally we obtain $n_e/n_c=0.52$ - for  $\lambda_0=1.47$~$\mu$m but only 0.16 - in the case of  $\lambda_0=0.8$~$\mu$m. 
More efficient energy transfer from the laser pulse to the electrons gives an advantage to the case with $\lambda_0=1.47$~$\mu$m. With a further increase in the ratio $n_e/n_c$ with increasing $\lambda_0$, the nonlinearity of the interaction of the plasma and electron currents becomes more significant, which, in particular, is expressed in a longer attenuation of the electric field components compared to a vacuum and even the appearance of secondary peaks in the absolute maxima of $E_x$, see Fig.~\ref{absMax25}(c). 
As for further increase of $a_0$, which should lead to higher kinetic energies, there is a risk of encountering beam filamentation during its propagation between the mirror and the focus, although the B-integral for propagation in neutral gas, estimated in the same way as in our previous article\cite{Lytova25}, is $\approx16.5\cdot(1.8/\lambda_0[\mu \text{m}])$ mrad, which is still small with respect to $2\pi/10$ rad. A more accurate modeling of nonlinear effects on beam propagation in this configuration will be the subject of future investigations. 

\subsection{Generation of electron beams in low-density air}

\begin{figure}[t]
	\centering{
		\includegraphics[scale=1.1,trim={5 15 20 25},clip]{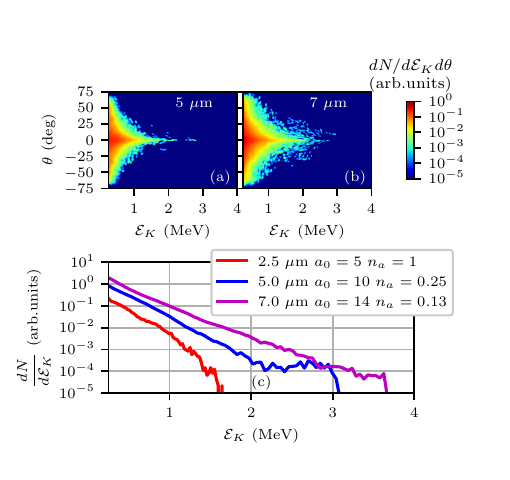}}
	\caption{Upper row: Energy-angular spectra of electrons accelerated forward by a tightly focused LP (in direction $x$) laser pulse with (a) $a_0 = 10$, $\lambda_0=5$ $\mu$m, $n_a=0.25$ and (b)  $a_0 = 14$, $\lambda_0=7$ $\mu$m, $n_a=0.125$ are shown at time $t = 91.5\cdot\lambda_0[\mu\text{m}]$ fs in $yz$ plane; (c) Angle-averaged energy spectra for the same parameters as for (a) and (b), and for comparison the case with $a_0 = 5$, $\lambda_0=2.5$ $\mu$m, $n_a=1$ is shown.}  \label{spec_5_7lp}
\end{figure}

In this section we demonstrate another way to increase the maximum kinetic energy of electrons, this time using longer wavelengths and lower gas densities. From previous simulations we know that at $a_0=5$ the maximum kinetic energy of the electron beam can be achieved for wavelength $\lambda_0=2.5$~$\mu$m.
As we discussed in the previous section and showed with previous simulations\cite{Lytova25}, simply increasing $a_0$ does not work - we will increase the ionization and get an overdense plasma where the pulse will be rapidly decaying. Thus, we need to change several parameters at once. Let us introduce two combinations that control the properties of the electron beam:
\begin{align}
	\Lambda_c(n_a,\lambda_0) &= n_a\lambda_0^2,\\
	\Lambda_i(a_0,\lambda_0) &= a_0^2\lambda_0^{-2},
\end{align}
where $n_a$ is the ratio of air density under given conditions and under standard conditions (above we considered only cases with $n_a=1$). Both $\Lambda$'s are responsible for the rate at which the plasma density approaches its critical value: $\Lambda_c$ is proportional to the ratio $n_a/n_c$ in the case of single ionization, while $\Lambda_i$ is proportional to the peak intensity, which determines the number of ionization electrons per atom. We know that parameters $\lambda_0=2.5$ $\mu$m and $a_0=5$ are optimal under standard conditions, when $n_a=1$, so we can introduce the following ``optimal'' coefficients
\begin{align}
	\Lambda_c^\text{o}&=\Lambda_c(1,2.5 \mu\text{m})=6.25\: \mu\text{m}^2,\\
	\Lambda_i^\text{o}&=\Lambda_i(5,2.5 \mu\text{m}) = 4 \: \mu\text{m}^{-2},
\end{align}
to use them to determine optimal conditions for different $n_a$. For example, if we want to accelerate electrons with laser strength $a_0=10$, we can do this using a laser with a wavelength $\lambda_0=a_0/\sqrt{\Lambda_i^\text{o}}=5$~$\mu$m, in a gas with an initial neutral density $n_a=\Lambda_c^\text{o}/\lambda_0^2=0.25$. The electron spectra simulated using these parameters are shown in Fig.~\ref{spec_5_7lp}(a),(c). We can proceed in the same way, increasing $a_0$ and $\lambda_0$ with a corresponding decrease in $n_a$, see Fig.~\ref{spec_5_7lp}(b),(c). As can be seen from the simulation results, we increase not only the maximum kinetic energy of electrons with this approach, but also their total number. In addition, the increase in $a_0$ occurs due to the increase in $\lambda_0$, while the intensity in W/cm$^2$ for all cases remains the same.

\begin{figure}[t]
	\centering{
		\includegraphics[scale=1,trim={0 0 0 0},clip]{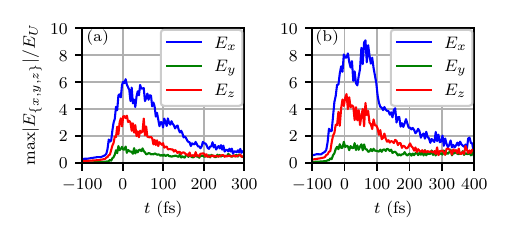}}
	\caption{Maximum over the entire simulation box values of the electric field components in time for LP pulse with (a) $\lambda_0=5$ $\mu$m, $a_0 = 10$, $n_a=0.25$ and (b) $\lambda_0=7$ $\mu$m, $a_0 = 14$, $n_a\approx0.13$.}  \label{absMax5_7}
\end{figure}

Note that for scaling we fixed $n_e/n_c=0.8$, so that in Figures~\ref{absMax5_7}(a),(b) we again see the same two-peak structure - a sign of nonlinear interaction between plasma and laser. Also, for all three cases shown in Fig.~\ref{spec_5_7lp} (c), approximately the same share of EM energy is spent on ionization and heating of the plasma, $\approx30-32\%$, despite the rather different maximum energies of the directed electron beams.

\section{Generation of electron beams in noble gases}
\label{sec:gases}
Other suitable media for generating directed beams of accelerated electrons are noble gases such as Ne, Ar, Kr. At laser intensities considered, in the range $10^{19}-10^{20}$~W/cm$^2$, one can expect ionization of the following gaseous media to the states shown in the table~\ref{ion_st}. Thus, we can follow the same procedure described above in section~\ref{ambient_air}: use a relatively small $\lambda_0$, say 1.47~$\mu$m, and increase $a_0$ to increase the maximum kinetic energy of the electrons.
\begin{table}[h]
	\caption{\label{ion_st} Molar ionization energies for upper ionization states expected under experimental conditions}
\begin{ruledtabular}
	\begin{tabular}{cccccc}
		Ion& N$^{5+}$  & O$^{6+}$ & Ne$^{6+}$ & Ar$^{8+}$ & Kr$^{8+}$ \\ \hline
		\makecell{Ionization \\ energy  (eV)}& 97.9& 138.1& 157.9 & 143.5 & 125.8  \\ 
	\end{tabular}
\end{ruledtabular}
\end{table}

\begin{figure}[h]
	\centering{
		\includegraphics[scale=1.1,trim={5 15 20 25},clip]{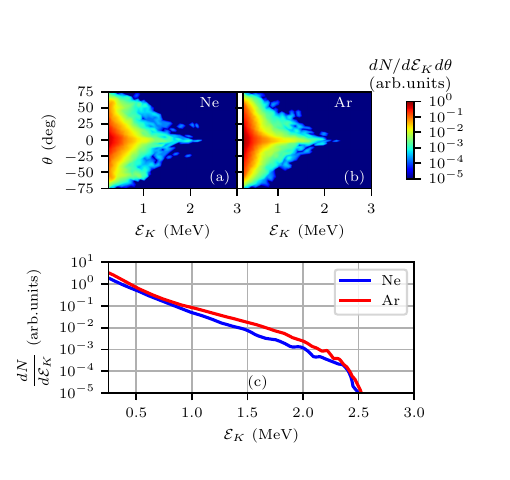}}
	\caption{Upper row: Energy-angular spectra ($yz$-plane) of electrons accelerated forward by a tightly focused LP (in direction $x$) laser pulse with $\lambda_0=1.47$ $\mu$m and (a) $a_0 = 10$ in Ne (b) $a_0 = 15$ in Ar at time $t = 91.5\cdot\lambda_0[\mu\text{m}]$ fs. (c) Angle-averaged energy spectrum for the same parameters as for (a),(b).}  \label{spec_Ne_Ar}
\end{figure}

Figure~\ref{spec_Ne_Ar} shows the spectra of electrons accelerated in Ne (a) and Ar (b) plasma. For the simulations shown, the average ionization states near the focus are $Z_\text{Ne}=4.4$ for $a_0=10$ and $Z_\text{Ar}=7.5$ for $a_0=15$. In these cases, the maximum values $n_e/n_c$ ratio near the focus are $0.46$ and $0.78$, respectively. As a consequence, relative field attenuation is stronger in Ar than in Ne. Namely, in Ne $E_x^\mathrm{max}= 7$, while in Ar - 8. As a result, we see practically the same maximum kinetic energies, despite essentially different laser setups for $a_0$ when focusing in vacuum.
As follows from Tab.~\ref{ion_st}, apparently, a similar picture should be observed with Kr, the ionization of which, with the same laser parameters $a_0=15$ and $\lambda_0=1.47$ $\mu$m, will lead  $Z_\text{Kr}=8$. In general, although noble gases can also be used as an acceleration medium, the maximum electron energies in the simulation are lower for the same laser parameters than in the case of focusing in ambient air - as can be seen, for example, from a comparison of Fig.~\ref{spec_a0_15} and Fig.~\ref{spec_Ne_Ar}, so this direction does not look particularly promising for practical applications..

\section{Conclusion}
\label{sec:conclusion}
Our study was motivated by recent and planned experiments on accelerating electron beams by tightly focusing an ultrashort laser pulse in gases. We considered several parameters of the laser pulse: polarization, field strength and wavelength as well as the number density of the gas medium. Comparing the results of the simulations performed for the LP, CP and RP pulses, we come to the conclusion that the LP pulse is preferable for ensuring the directionality of the accelerated electron beam. To increase the energy and number of particles in the electron beam under ambient air conditions, we found that the most optimal wavelength is about 1.5 $\mu$m and $a_0\geq10$. Unfortunately, we have not found simple scaling laws for selecting laser parameters to accelerate electrons to predetermined energies as all the parameters of the system are nonlinearly related. In addition to direct acceleration by relativistic PMF, nonlinear interactions of the laser with the plasma can play an important role in the near-critical regime. Although this can contribute to electron acceleration, this process is difficult to control and requires a fine tuning of the laser and gas parameters. At the same time, we need to keep in mind that with increasing $a_0$, the risk of pulse filamentation increases due to the nonlinearity of propagation before focusing, which should become the topic of future research. 

We also found a potentially useful scaling that increases the maximum energy of accelerated electrons in thin air when using lasers with wavelength > 2.5 $\mu$m (the value at which the maximum is achieved in ambient air). This follows from two simple assumptions of keeping the degree of ionization and acceleration efficiency the same as when obtaining the maximum energy in ambient air.

Finally we demonstrated that some noble gases with the same number density as the ambient air can be suitable media for accelerating electrons in a tightly focused laser fields. However, here it is necessary to pay attention to the proximity of the electron density to the critical one, maintaining a balance between the number of electrons in the beam and the attenuation of the laser pulse in the near-critical plasma. So far we do not see any advantages of these gas environments over the ambient air. Finally, we found that the coordinated change in the density of the gas medium, the laser wavelength and its strength opens up another opportunity to improve the properties of directed electron beams.

\appendix
\section{Exact solution for ultrashort CP tightly-focused pulse in vacuum}\label{exact_CP}

To obtain the solution, we use the approach developed by April \cite{April_2010}, as we did previously for the LP pulse\cite{Lytova25}, where the whole derivation procedure was described in detail. So here we only present the difference in the initial expressions for the Hertz potentials and present the final result. 
Within this approach, the EM field components are expressed in terms of the electric ${\bf \Pi}_{\mathrm{E}}$ and magnetic ${\bf \Pi}_{\mathrm{B}}$ Hertz potentials. Here, we use the convention that ${\bf \Pi}_{\mathrm{B}} = \mu_{0}{\bf \Pi}_{\mathrm{m}}$ and ${\bf \Pi}_{\mathrm{E}} = {\bf \Pi}_{\mathrm{e}}/\epsilon_{0}$, while ${\bf \Pi}_{\mathrm{m},\mathrm{e}}$ are used by other authors \cite{Jackson:100964}. For a tightly focused CP pulse propagating in the $z$-direction and polarized in the $xy$-plane, the Hertz potentials are
\begin{align}
	\label{cp_pol_e}
	{\bf\Pi}_{\mathrm{e}}&={\bf e}_x\Psi+i\:{\bf e}_y\Psi,\\
	\label{cp_pol_m}
	{\bf\Pi}_{\mathrm{m}}&=c^{-1}\{-i\:{\bf e}_x\Psi+{\bf e}_y\Psi\},
\end{align}
where $\Psi(x,y,z,t)$ is a scalar function, to be determined from some additional conditions. Thus the following solution can be obtained in closed form for isodiffracting pulse with Poisson-like spectrum: 
\begin{align}\label{Ex}
	E_x =\dfrac{\Psi_0}{b^3}\Big(\dfrac{x(x+iy)}{r^2}\mathcal{K}_1- \mathcal{K}_2\Big),
\end{align}
\begin{align}
	E_y = \dfrac{\Psi_0}{b^3}\Big( \dfrac{y(x+iy)}{r^2}\mathcal{K}_1-i\mathcal{K}_2\Big),
\end{align}
\begin{align}
	E_z=\dfrac{\Psi_0}{b^3}\dfrac{(x+iy)}{r}\mathcal{K}_3,
\end{align}
\begin{align}
	cB_x = \dfrac{\Psi_0}{b^3}\Big( \dfrac{x(y-ix)}{r^2}\mathcal{K}_1 + i\mathcal{K}_2\Big),
\end{align}
\begin{align}
	cB_y = \dfrac{\Psi_0}{b^3}\Big(\dfrac{y(y-ix)}{r^2} \mathcal{K}_1-\mathcal{K}_2\Big),
\end{align}
\begin{align}\label{Bz}
	cB_z = \dfrac{\Psi_0}{b^3}\dfrac{(y-ix)}{r}\mathcal{K}_3,
\end{align}
where the complex-valued functions are defined as:
\begin{align}\label{K1}
	\mathcal{K}_1&=\dfrac{\sin^2\tilde{\theta}}{\widetilde{\mathcal{R}}}\Big(\dfrac{3\mathcal{G}^{(0)}_-}{\widetilde{\mathcal{R}}^2}-\dfrac{3\mathcal{G}^{(1)}_+}{\widetilde{\mathcal{R}}}+ \mathcal{G}^{(2)}_- \Big),
\end{align}
\begin{align}
	\nonumber\mathcal{K}_2=\dfrac{1}{\widetilde{\mathcal{R}}}\Big(\dfrac{\mathcal{G}^{(0)}_-}{\widetilde{\mathcal{R}}^2}+&\dfrac{\mathcal{G}^{(1)}_-\cos\tilde{\theta}}{\widetilde{\mathcal{R}}} -
	\dfrac{\mathcal{G}^{(1)}_+}{\widetilde{\mathcal{R}}}+\\
	&\mathcal{G}^{(2)}_--
	\mathcal{G}^{(2)}_+\cos\tilde{\theta}\Big),
\end{align}
\begin{align}\label{K3}
	\nonumber\mathcal{K}_3=\dfrac{\sin\tilde{\theta}}{\widetilde{\mathcal{R}}}\Big(\dfrac{3\mathcal{G}^{(0)}_-\cos\tilde{\theta}}{\widetilde{\mathcal{R}}^2}+&\dfrac{\mathcal{G}^{(1)}_-}{\widetilde{\mathcal{R}}}-\dfrac{3\mathcal{G}^{(1)}_+\cos\tilde{\theta}}{\widetilde{\mathcal{R}}}+\\
	&\mathcal{G}^{(2)}_-\cos\tilde{\theta}- \mathcal{G}^{(2)}_+ \Big).
\end{align}
In \eqref{K1}-\eqref{K3} the functions
\begin{equation}
	\mathcal{G}_\pm^{(n)}=\Big(\dfrac{b}{c}\Big)^nG_\pm^{(n)} =g^{(n)}(\tilde{\tau}_+)\pm g^{(n)}(\tilde{\tau}_-),
\end{equation} 
are expressed through functions
\begin{equation}\label{poisson}
	g^{(n)}(\tau)=e^{-i\phi_0}\dfrac{\Gamma(s+n+1)}{\Gamma(s+1)}\Big(\dfrac{ikb}{s}\Big)^n\Big(1-\dfrac{i\tau}{s}\Big)^{-(s+n+1)}.
\end{equation}
with arguments
\begin{equation}
	\tau_\pm = \tau + k_0b(\pm\widetilde{\mathcal{R}}+i),
\end{equation}
where $\tau = \omega_0t$, $\widetilde{\mathcal{R}} = \sqrt{\rho^2+(\zeta + i)^2}$, $\rho= \sqrt{x^2+y^2}/b$, $\zeta = z/b$, and $\phi_0$ is a constant phase. Also in the expressions for $\mathcal{K}_{1}-\mathcal{K}_3$ we use the notation: $\sin\tilde{\theta}=\dfrac{\rho}{\widetilde{\mathcal{R}}}$, $\cos\tilde{\theta} = \dfrac{\zeta + i}{\widetilde{\mathcal{R}}}$. Note that in the case of a finite beam, the solution is approximately CP, being purely CP only on the optical axis \cite{John_Lekner_2003}. This is also observed in the obtained solution. 

The above solution depends on three arguments, which allows us to relate it to the laser parameters in the experiment:
\begin{enumerate}
	\item Confocal parameter $b$ expressed through the numerical aperture NA
	\begin{align}\label{b_NA}
		b = \frac{2\sqrt{1-\text{NA}^2}}{k_0\text{NA}^2}.
	\end{align}
	\item Parameter $s$  determines the duration of the ultrashort pulse in time \cite{Caron}
	\begin{align}\label{tau_w0}
		\tau_{\text{FWHM}}=\sqrt{2}s\sqrt{2^{2/(s+1)}-1}/\omega_0.
	\end{align}
	\item Amplitude $\Psi_0$ depends on the laser strength parameter $a_0$ and two above parameters
	\begin{equation}\label{psi0}
		\Psi_0 = -\frac{a_0k_0b^3m_ec^2}{e\mathrm{Re}\{\mathcal{K}_{2}^{\mathrm{foc}}\}},
	\end{equation}
	where $e$ in the denominator is the elementary charge and
	\begin{align}\label{K2_max}
		\nonumber\mathcal{K}_{2}^{\mathrm{foc}}&=ie^{-i\phi_0}\Big[(1+2k_0b/s)^{-(s+1)}-1+\\
		&2k_0b(s+1)/s-2k_0^2b^2(s+2)(s+1)/s^2 \Big],
	\end{align}
	so for the peak intensity we take $\phi_0=\pi/2$.
\end{enumerate}

Note that all relations \eqref{b_NA}-\eqref{K2_max}  also hold for LP pulses \cite{Lytova25}, but only \eqref{b_NA} and \eqref{tau_w0} are valid for the RP case. For the RP solutions  \cite{Marceau:12} instead of \eqref{psi0} and \eqref{K2_max}, we should use
\begin{equation}\label{psi0_RP}
	\Psi_0^\text{RP} = -\frac{a_0k_0b^3m_ec^2}{2e\mathrm{Re}\{\mathcal{K}_\text{RP}^{\mathrm{foc}}\}},
\end{equation}
where 
\begin{align}\label{K2_max_RP}
	\nonumber\mathcal{K}_\text{RP}^{\mathrm{foc}}&=ie^{-i\phi_0}\Big[1-(1+2k_0b/s)^{-(s+1)}-\\
	&k_0b(s+1)/s\big(1+(1+2k_0b/s)^{-(s+2)}\big)\Big]
\end{align}
now refers to the component $E_z$.

For comparison with experiment, it may also be useful to relate the amplitude $\Psi_0$ to the laser energy per pulse $\mathcal{E}_L$:
\begin{align}
	\Psi_0=b^2\sqrt{\frac{\mathcal{E}_L}{\pi\varepsilon_0\mathcal{Z}(s,k_0b)}},
\end{align}
where in the case of LP pulse
\begin{align}
\nonumber\mathcal{Z}^{LP}(s.k_0b)&=\int_{-R}^{R}\Big[\int_0^{\sqrt{R^2-z^2}/b}\Big((\mathrm{Re}\{\mathcal{K}_1^0-\mathcal{K}_2^0\})^2\\
&+(\mathrm{Re}\{\mathcal{K}_2^0\})^2+(\mathrm{Re}\{\mathcal{K}_3^0\})^2 \Big)\rho d\rho\Big] dz,
\end{align}
while for CP pulse
\begin{align}
	\nonumber\mathcal{Z}^{CP}(s.k_0b)&=\nonumber\mathcal{Z}^{LP}\!+\!\int_{-R}^{R}\!\Big[\int_0^{\sqrt{R^2-z^2}/b}\!\Big((\mathrm{Im}\{\mathcal{K}_1^0\!-\!\mathcal{K}_2^0\})^2\\
	&+(\mathrm{Im}\{\mathcal{K}_2^0\})^2+(\mathrm{Im}\{\mathcal{K}_3^0\})^2 \Big)\rho d\rho\Big] dz,
\end{align}
with functions $\mathcal{K}_{1,2,3}^0 \equiv\mathcal{K}_{1,2,3}(r,z,0)$ taken at $t=0$, when the pulse is most compact. The latter circumstance also allows, when calculating the above integrals, to choose a smaller value of the radius $R$ of the region within which we are calculating the pulse energy.

\begin{acknowledgments}
This research was enabled in part by support provided by the Digital Research Alliance of Canada  \url{alliancecan.ca.}
\end{acknowledgments}

% If in two-column mode, this environment will change to single-column format so that long equations can be displayed. 
% Use only when necessary.
%\begin{widetext}
%$$\mbox{put long equation here}$$
%\end{widetext}

% Figures should be put into the text as floats. 
% Use the graphics or graphicx packages (distributed with LaTeX2e).
% See the LaTeX Graphics Companion by Michel Goosens, Sebastian Rahtz, and Frank Mittelbach for examples. 
%
% Here is an example of the general form of a figure:
% Fill in the caption in the braces of the \caption{} command. 
% Put the label that you will use with \ref{} command in the braces of the \label{} command.
%
% \begin{figure}
% \includegraphics{}%
% \caption{\label{}}%
% \end{figure}

% Tables may be be put in the text as floats.
% Here is an example of the general form of a table:
% Fill in the caption in the braces of the \caption{} command. Put the label
% that you will use with \ref{} command in the braces of the \label{} command.
% Insert the column specifiers (l, r, c, d, etc.) in the empty braces of the
% \begin{tabular}{} command.
%
% \begin{table}
% \caption{\label{} }
% \begin{tabular}{}
% \end{tabular}
% \end{table}

% If you have acknowledgments, this puts in the proper section head.
%\begin{acknowledgments}
% Put your acknowledgments here.
%\end{acknowledgments}

% Create the reference section using BibTeX:
\bibliography{mybibfile}

%aipnum4-2.bst 2019-01-14 (MD) hand-edited version of apsrev4-1.bst
%Control: key (0)
%Control: author (8) initials jnrlst
%Control: editor formatted (1) identically to author
%Control: production of article title (0) allowed
%Control: page (1) range
%Control: year (1) truncated
%Control: production of eprint (0) enabled
\providecommand{\noopsort}[1]{}\providecommand{\singleletter}[1]{#1}%
\begin{thebibliography}{49}%
\makeatletter
\providecommand \@ifxundefined [1]{%
 \@ifx{#1\undefined}
}%
\providecommand \@ifnum [1]{%
 \ifnum #1\expandafter \@firstoftwo
 \else \expandafter \@secondoftwo
 \fi
}%
\providecommand \@ifx [1]{%
 \ifx #1\expandafter \@firstoftwo
 \else \expandafter \@secondoftwo
 \fi
}%
\providecommand \natexlab [1]{#1}%
\providecommand \enquote  [1]{``#1''}%
\providecommand \bibnamefont  [1]{#1}%
\providecommand \bibfnamefont [1]{#1}%
\providecommand \citenamefont [1]{#1}%
\providecommand \href@noop [0]{\@secondoftwo}%
\providecommand \href [0]{\begingroup \@sanitize@url \@href}%
\providecommand \@href[1]{\@@startlink{#1}\@@href}%
\providecommand \@@href[1]{\endgroup#1\@@endlink}%
\providecommand \@sanitize@url [0]{\catcode `\\12\catcode `\$12\catcode
  `\&12\catcode `\#12\catcode `\^12\catcode `\_12\catcode `\%12\relax}%
\providecommand \@@startlink[1]{}%
\providecommand \@@endlink[0]{}%
\providecommand \url  [0]{\begingroup\@sanitize@url \@url }%
\providecommand \@url [1]{\endgroup\@href {#1}{\urlprefix }}%
\providecommand \urlprefix  [0]{URL }%
\providecommand \Eprint [0]{\href }%
\providecommand \doibase [0]{https://doi.org/}%
\providecommand \selectlanguage [0]{\@gobble}%
\providecommand \bibinfo  [0]{\@secondoftwo}%
\providecommand \bibfield  [0]{\@secondoftwo}%
\providecommand \translation [1]{[#1]}%
\providecommand \BibitemOpen [0]{}%
\providecommand \bibitemStop [0]{}%
\providecommand \bibitemNoStop [0]{.\EOS\space}%
\providecommand \EOS [0]{\spacefactor3000\relax}%
\providecommand \BibitemShut  [1]{\csname bibitem#1\endcsname}%
\let\auto@bib@innerbib\@empty
%</preamble>
\bibitem [{\citenamefont {Vallières}\ \emph {et~al.}(2024)\citenamefont
  {Vallières}, \citenamefont {Powell}, \citenamefont {Connell}, \citenamefont
  {Evans}, \citenamefont {Lytova}, \citenamefont {Fillion-Gourdeau},
  \citenamefont {Fourmaux}, \citenamefont {Payeur}, \citenamefont {Lassonde},
  \citenamefont {MacLean},\ and\ \citenamefont {Légaré}}]{vallieres2022}%
  \BibitemOpen
  \bibfield  {author} {\bibinfo {author} {\bibfnamefont {S.}~\bibnamefont
  {Vallières}}, \bibinfo {author} {\bibfnamefont {J.}~\bibnamefont {Powell}},
  \bibinfo {author} {\bibfnamefont {T.}~\bibnamefont {Connell}}, \bibinfo
  {author} {\bibfnamefont {M.}~\bibnamefont {Evans}}, \bibinfo {author}
  {\bibfnamefont {M.}~\bibnamefont {Lytova}}, \bibinfo {author} {\bibfnamefont
  {F.}~\bibnamefont {Fillion-Gourdeau}}, \bibinfo {author} {\bibfnamefont
  {S.}~\bibnamefont {Fourmaux}}, \bibinfo {author} {\bibfnamefont
  {S.}~\bibnamefont {Payeur}}, \bibinfo {author} {\bibfnamefont
  {P.}~\bibnamefont {Lassonde}}, \bibinfo {author} {\bibfnamefont
  {S.}~\bibnamefont {MacLean}},\ and\ \bibinfo {author} {\bibfnamefont
  {F.}~\bibnamefont {Légaré}},\ }\bibfield  {title} {\enquote {\bibinfo
  {title} {High dose-rate mev electron beam from a tightly-focused femtosecond
  ir laser in ambient air},}\ }\href
  {https://doi.org/https://doi.org/10.1002/lpor.202300078} {\bibfield
  {journal} {\bibinfo  {journal} {Laser \& Photonics Reviews}\ }\textbf
  {\bibinfo {volume} {18}},\ \bibinfo {pages} {2300078} (\bibinfo {year}
  {2024})}\BibitemShut {NoStop}%
\bibitem [{\citenamefont {Lytova}\ \emph {et~al.}(2025)\citenamefont {Lytova},
  \citenamefont {Fillion-Gourdeau}, \citenamefont {Valli\`eres}, \citenamefont
  {Fourmaux}, \citenamefont {Payeur}, \citenamefont {Powell}, \citenamefont
  {L\'egar\'e},\ and\ \citenamefont {MacLean}}]{Lytova25}%
  \BibitemOpen
  \bibfield  {author} {\bibinfo {author} {\bibfnamefont {M.}~\bibnamefont
  {Lytova}}, \bibinfo {author} {\bibfnamefont {F.}~\bibnamefont
  {Fillion-Gourdeau}}, \bibinfo {author} {\bibfnamefont {S.}~\bibnamefont
  {Valli\`eres}}, \bibinfo {author} {\bibfnamefont {S.}~\bibnamefont
  {Fourmaux}}, \bibinfo {author} {\bibfnamefont {S.}~\bibnamefont {Payeur}},
  \bibinfo {author} {\bibfnamefont {J.}~\bibnamefont {Powell}}, \bibinfo
  {author} {\bibfnamefont {F.}~\bibnamefont {L\'egar\'e}},\ and\ \bibinfo
  {author} {\bibfnamefont {S.}~\bibnamefont {MacLean}},\ }\bibfield  {title}
  {\enquote {\bibinfo {title} {Electron acceleration in ambient air using
  tightly focused ultrashort infrared laser beams},}\ }\href
  {https://doi.org/10.1103/PhysRevE.111.035210} {\bibfield  {journal} {\bibinfo
   {journal} {Phys. Rev. E}\ }\textbf {\bibinfo {volume} {111}},\ \bibinfo
  {pages} {035210} (\bibinfo {year} {2025})}\BibitemShut {NoStop}%
\bibitem [{\citenamefont {Vozenin}\ \emph {et~al.}(2024)\citenamefont
  {Vozenin}, \citenamefont {Loo}, \citenamefont {Tantawi}, \citenamefont
  {Maxim}, \citenamefont {Spitz}, \citenamefont {Bailat},\ and\ \citenamefont
  {Limoli}}]{Vozenin}%
  \BibitemOpen
  \bibfield  {author} {\bibinfo {author} {\bibfnamefont {M.-C.}\ \bibnamefont
  {Vozenin}}, \bibinfo {author} {\bibfnamefont {B.~W.}\ \bibnamefont {Loo}},
  \bibinfo {author} {\bibfnamefont {S.}~\bibnamefont {Tantawi}}, \bibinfo
  {author} {\bibfnamefont {P.~G.}\ \bibnamefont {Maxim}}, \bibinfo {author}
  {\bibfnamefont {D.~R.}\ \bibnamefont {Spitz}}, \bibinfo {author}
  {\bibfnamefont {C.}~\bibnamefont {Bailat}},\ and\ \bibinfo {author}
  {\bibfnamefont {C.~L.}\ \bibnamefont {Limoli}},\ }\bibfield  {title}
  {\enquote {\bibinfo {title} {Flash: New intersection of physics, chemistry,
  biology, and cancer medicine},}\ }\href
  {https://doi.org/10.1103/RevModPhys.96.035002} {\bibfield  {journal}
  {\bibinfo  {journal} {Rev. Mod. Phys.}\ }\textbf {\bibinfo {volume} {96}},\
  \bibinfo {pages} {035002} (\bibinfo {year} {2024})}\BibitemShut {NoStop}%
\bibitem [{\citenamefont {Cleland}(2006)}]{Cleland:1005393}%
  \BibitemOpen
  \bibfield  {author} {\bibinfo {author} {\bibfnamefont {M.~R.}\ \bibnamefont
  {Cleland}},\ }\bibfield  {title} {\enquote {\bibinfo {title} {{Industrial
  applications of electron accelerators}},}\ }\href
  {https://doi.org/10.5170/CERN-2006-012.383} {\  (\bibinfo {year} {2006}),\
  10.5170/CERN-2006-012.383}\BibitemShut {NoStop}%
\bibitem [{\citenamefont {Hamm}\ and\ \citenamefont
  {Hamm}(2012)}]{hamm2012industrial}%
  \BibitemOpen
  \bibfield  {author} {\bibinfo {author} {\bibfnamefont {R.~W.}\ \bibnamefont
  {Hamm}}\ and\ \bibinfo {author} {\bibfnamefont {M.~E.}\ \bibnamefont
  {Hamm}},\ }\bibfield  {title} {\enquote {\bibinfo {title} {Industrial
  accelerators and their applications},}\ }\href@noop {} {\  (\bibinfo {year}
  {2012})}\BibitemShut {NoStop}%
\bibitem [{\citenamefont {Daido}, \citenamefont {Nishiuchi},\ and\
  \citenamefont {Pirozhkov}(2012)}]{Daido_2012}%
  \BibitemOpen
  \bibfield  {author} {\bibinfo {author} {\bibfnamefont {H.}~\bibnamefont
  {Daido}}, \bibinfo {author} {\bibfnamefont {M.}~\bibnamefont {Nishiuchi}},\
  and\ \bibinfo {author} {\bibfnamefont {A.~S.}\ \bibnamefont {Pirozhkov}},\
  }\bibfield  {title} {\enquote {\bibinfo {title} {Review of laser-driven ion
  sources and their applications},}\ }\href
  {https://doi.org/10.1088/0034-4885/75/5/056401} {\bibfield  {journal}
  {\bibinfo  {journal} {Reports on Progress in Physics}\ }\textbf {\bibinfo
  {volume} {75}},\ \bibinfo {pages} {056401} (\bibinfo {year}
  {2012})}\BibitemShut {NoStop}%
\bibitem [{\citenamefont {Esarey}, \citenamefont {Schroeder},\ and\
  \citenamefont {Leemans}(2009)}]{RevModPhys.81.1229}%
  \BibitemOpen
  \bibfield  {author} {\bibinfo {author} {\bibfnamefont {E.}~\bibnamefont
  {Esarey}}, \bibinfo {author} {\bibfnamefont {C.~B.}\ \bibnamefont
  {Schroeder}},\ and\ \bibinfo {author} {\bibfnamefont {W.~P.}\ \bibnamefont
  {Leemans}},\ }\bibfield  {title} {\enquote {\bibinfo {title} {Physics of
  laser-driven plasma-based electron accelerators},}\ }\href
  {https://doi.org/10.1103/RevModPhys.81.1229} {\bibfield  {journal} {\bibinfo
  {journal} {Rev. Mod. Phys.}\ }\textbf {\bibinfo {volume} {81}},\ \bibinfo
  {pages} {1229--1285} (\bibinfo {year} {2009})}\BibitemShut {NoStop}%
\bibitem [{\citenamefont {Tajima}\ and\ \citenamefont {Dawson}(1979)}]{Tajima}%
  \BibitemOpen
  \bibfield  {author} {\bibinfo {author} {\bibfnamefont {T.}~\bibnamefont
  {Tajima}}\ and\ \bibinfo {author} {\bibfnamefont {J.~M.}\ \bibnamefont
  {Dawson}},\ }\bibfield  {title} {\enquote {\bibinfo {title} {Laser electron
  accelerator},}\ }\href {https://doi.org/10.1103/PhysRevLett.43.267}
  {\bibfield  {journal} {\bibinfo  {journal} {Phys. Rev. Lett.}\ }\textbf
  {\bibinfo {volume} {43}},\ \bibinfo {pages} {267--270} (\bibinfo {year}
  {1979})}\BibitemShut {NoStop}%
\bibitem [{\citenamefont {Gu{\'e}not}\ \emph {et~al.}(2017)\citenamefont
  {Gu{\'e}not}, \citenamefont {Gustas}, \citenamefont {Vernier}, \citenamefont
  {Beaurepaire}, \citenamefont {B{\"o}hle}, \citenamefont {Bocoum},
  \citenamefont {Lozano}, \citenamefont {Jullien}, \citenamefont
  {Lopez-Martens}, \citenamefont {Lifschitz} \emph
  {et~al.}}]{guenot2017relativistic}%
  \BibitemOpen
  \bibfield  {author} {\bibinfo {author} {\bibfnamefont {D.}~\bibnamefont
  {Gu{\'e}not}}, \bibinfo {author} {\bibfnamefont {D.}~\bibnamefont {Gustas}},
  \bibinfo {author} {\bibfnamefont {A.}~\bibnamefont {Vernier}}, \bibinfo
  {author} {\bibfnamefont {B.}~\bibnamefont {Beaurepaire}}, \bibinfo {author}
  {\bibfnamefont {F.}~\bibnamefont {B{\"o}hle}}, \bibinfo {author}
  {\bibfnamefont {M.}~\bibnamefont {Bocoum}}, \bibinfo {author} {\bibfnamefont
  {M.}~\bibnamefont {Lozano}}, \bibinfo {author} {\bibfnamefont
  {A.}~\bibnamefont {Jullien}}, \bibinfo {author} {\bibfnamefont
  {R.}~\bibnamefont {Lopez-Martens}}, \bibinfo {author} {\bibfnamefont
  {A.}~\bibnamefont {Lifschitz}}, \emph {et~al.},\ }\bibfield  {title}
  {\enquote {\bibinfo {title} {Relativistic electron beams driven by khz
  single-cycle light pulses},}\ }\href
  {https://doi.org/https://doi.org/10.1038/nphoton.2017.46} {\bibfield
  {journal} {\bibinfo  {journal} {Nature Photonics}\ }\textbf {\bibinfo
  {volume} {11}},\ \bibinfo {pages} {293--296} (\bibinfo {year}
  {2017})}\BibitemShut {NoStop}%
\bibitem [{\citenamefont {Gustas}\ \emph {et~al.}(2018)\citenamefont {Gustas},
  \citenamefont {Gu\'enot}, \citenamefont {Vernier}, \citenamefont {Dutt},
  \citenamefont {B\"ohle}, \citenamefont {Lopez-Martens}, \citenamefont
  {Lifschitz},\ and\ \citenamefont {Faure}}]{PhysRevAccelBeams.21.013401}%
  \BibitemOpen
  \bibfield  {author} {\bibinfo {author} {\bibfnamefont {D.}~\bibnamefont
  {Gustas}}, \bibinfo {author} {\bibfnamefont {D.}~\bibnamefont {Gu\'enot}},
  \bibinfo {author} {\bibfnamefont {A.}~\bibnamefont {Vernier}}, \bibinfo
  {author} {\bibfnamefont {S.}~\bibnamefont {Dutt}}, \bibinfo {author}
  {\bibfnamefont {F.}~\bibnamefont {B\"ohle}}, \bibinfo {author} {\bibfnamefont
  {R.}~\bibnamefont {Lopez-Martens}}, \bibinfo {author} {\bibfnamefont
  {A.}~\bibnamefont {Lifschitz}},\ and\ \bibinfo {author} {\bibfnamefont
  {J.}~\bibnamefont {Faure}},\ }\bibfield  {title} {\enquote {\bibinfo {title}
  {High-charge relativistic electron bunches from a khz laser-plasma
  accelerator},}\ }\href {https://doi.org/10.1103/PhysRevAccelBeams.21.013401}
  {\bibfield  {journal} {\bibinfo  {journal} {Phys. Rev. Accel. Beams}\
  }\textbf {\bibinfo {volume} {21}},\ \bibinfo {pages} {013401} (\bibinfo
  {year} {2018})}\BibitemShut {NoStop}%
\bibitem [{\citenamefont {Lu}\ \emph {et~al.}(2007)\citenamefont {Lu},
  \citenamefont {Tzoufras}, \citenamefont {Joshi}, \citenamefont {Tsung},
  \citenamefont {Mori}, \citenamefont {Vieira}, \citenamefont {Fonseca},\ and\
  \citenamefont {Silva}}]{PhysRevSTAB.10.061301}%
  \BibitemOpen
  \bibfield  {author} {\bibinfo {author} {\bibfnamefont {W.}~\bibnamefont
  {Lu}}, \bibinfo {author} {\bibfnamefont {M.}~\bibnamefont {Tzoufras}},
  \bibinfo {author} {\bibfnamefont {C.}~\bibnamefont {Joshi}}, \bibinfo
  {author} {\bibfnamefont {F.~S.}\ \bibnamefont {Tsung}}, \bibinfo {author}
  {\bibfnamefont {W.~B.}\ \bibnamefont {Mori}}, \bibinfo {author}
  {\bibfnamefont {J.}~\bibnamefont {Vieira}}, \bibinfo {author} {\bibfnamefont
  {R.~A.}\ \bibnamefont {Fonseca}},\ and\ \bibinfo {author} {\bibfnamefont
  {L.~O.}\ \bibnamefont {Silva}},\ }\bibfield  {title} {\enquote {\bibinfo
  {title} {Generating multi-gev electron bunches using single stage laser
  wakefield acceleration in a 3d nonlinear regime},}\ }\href
  {https://doi.org/10.1103/PhysRevSTAB.10.061301} {\bibfield  {journal}
  {\bibinfo  {journal} {Phys. Rev. ST Accel. Beams}\ }\textbf {\bibinfo
  {volume} {10}},\ \bibinfo {pages} {061301} (\bibinfo {year}
  {2007})}\BibitemShut {NoStop}%
\bibitem [{\citenamefont {Nakamura}\ \emph {et~al.}(2007)\citenamefont
  {Nakamura}, \citenamefont {Nagler}, \citenamefont {Tóth}, \citenamefont
  {Geddes}, \citenamefont {Schroeder}, \citenamefont {Esarey}, \citenamefont
  {Leemans}, \citenamefont {Gonsalves},\ and\ \citenamefont
  {Hooker}}]{10.1063/1.2718524}%
  \BibitemOpen
  \bibfield  {author} {\bibinfo {author} {\bibfnamefont {K.}~\bibnamefont
  {Nakamura}}, \bibinfo {author} {\bibfnamefont {B.}~\bibnamefont {Nagler}},
  \bibinfo {author} {\bibfnamefont {C.}~\bibnamefont {Tóth}}, \bibinfo
  {author} {\bibfnamefont {C.~G.~R.}\ \bibnamefont {Geddes}}, \bibinfo {author}
  {\bibfnamefont {C.~B.}\ \bibnamefont {Schroeder}}, \bibinfo {author}
  {\bibfnamefont {E.}~\bibnamefont {Esarey}}, \bibinfo {author} {\bibfnamefont
  {W.~P.}\ \bibnamefont {Leemans}}, \bibinfo {author} {\bibfnamefont {A.~J.}\
  \bibnamefont {Gonsalves}},\ and\ \bibinfo {author} {\bibfnamefont {S.~M.}\
  \bibnamefont {Hooker}},\ }\bibfield  {title} {\enquote {\bibinfo {title} {Gev
  electron beams from a centimeter-scale channel guided laser wakefield
  acceleratora)},}\ }\href {https://doi.org/10.1063/1.2718524} {\bibfield
  {journal} {\bibinfo  {journal} {Physics of Plasmas}\ }\textbf {\bibinfo
  {volume} {14}},\ \bibinfo {pages} {056708} (\bibinfo {year}
  {2007})}\BibitemShut {NoStop}%
\bibitem [{\citenamefont {Clayton}\ \emph {et~al.}(2010)\citenamefont
  {Clayton}, \citenamefont {Ralph}, \citenamefont {Albert}, \citenamefont
  {Fonseca}, \citenamefont {Glenzer}, \citenamefont {Joshi}, \citenamefont
  {Lu}, \citenamefont {Marsh}, \citenamefont {Martins}, \citenamefont {Mori},
  \citenamefont {Pak}, \citenamefont {Tsung}, \citenamefont {Pollock},
  \citenamefont {Ross}, \citenamefont {Silva},\ and\ \citenamefont
  {Froula}}]{PhysRevLett.105.105003}%
  \BibitemOpen
  \bibfield  {author} {\bibinfo {author} {\bibfnamefont {C.~E.}\ \bibnamefont
  {Clayton}}, \bibinfo {author} {\bibfnamefont {J.~E.}\ \bibnamefont {Ralph}},
  \bibinfo {author} {\bibfnamefont {F.}~\bibnamefont {Albert}}, \bibinfo
  {author} {\bibfnamefont {R.~A.}\ \bibnamefont {Fonseca}}, \bibinfo {author}
  {\bibfnamefont {S.~H.}\ \bibnamefont {Glenzer}}, \bibinfo {author}
  {\bibfnamefont {C.}~\bibnamefont {Joshi}}, \bibinfo {author} {\bibfnamefont
  {W.}~\bibnamefont {Lu}}, \bibinfo {author} {\bibfnamefont {K.~A.}\
  \bibnamefont {Marsh}}, \bibinfo {author} {\bibfnamefont {S.~F.}\ \bibnamefont
  {Martins}}, \bibinfo {author} {\bibfnamefont {W.~B.}\ \bibnamefont {Mori}},
  \bibinfo {author} {\bibfnamefont {A.}~\bibnamefont {Pak}}, \bibinfo {author}
  {\bibfnamefont {F.~S.}\ \bibnamefont {Tsung}}, \bibinfo {author}
  {\bibfnamefont {B.~B.}\ \bibnamefont {Pollock}}, \bibinfo {author}
  {\bibfnamefont {J.~S.}\ \bibnamefont {Ross}}, \bibinfo {author}
  {\bibfnamefont {L.~O.}\ \bibnamefont {Silva}},\ and\ \bibinfo {author}
  {\bibfnamefont {D.~H.}\ \bibnamefont {Froula}},\ }\bibfield  {title}
  {\enquote {\bibinfo {title} {Self-guided laser wakefield acceleration beyond
  1 gev using ionization-induced injection},}\ }\href
  {https://doi.org/10.1103/PhysRevLett.105.105003} {\bibfield  {journal}
  {\bibinfo  {journal} {Phys. Rev. Lett.}\ }\textbf {\bibinfo {volume} {105}},\
  \bibinfo {pages} {105003} (\bibinfo {year} {2010})}\BibitemShut {NoStop}%
\bibitem [{\citenamefont {Miao}\ \emph {et~al.}(2022)\citenamefont {Miao},
  \citenamefont {Shrock}, \citenamefont {Feder}, \citenamefont {Hollinger},
  \citenamefont {Morrison}, \citenamefont {Nedbailo}, \citenamefont {Picksley},
  \citenamefont {Song}, \citenamefont {Wang}, \citenamefont {Rocca},\ and\
  \citenamefont {Milchberg}}]{PhysRevX.12.031038}%
  \BibitemOpen
  \bibfield  {author} {\bibinfo {author} {\bibfnamefont {B.}~\bibnamefont
  {Miao}}, \bibinfo {author} {\bibfnamefont {J.~E.}\ \bibnamefont {Shrock}},
  \bibinfo {author} {\bibfnamefont {L.}~\bibnamefont {Feder}}, \bibinfo
  {author} {\bibfnamefont {R.~C.}\ \bibnamefont {Hollinger}}, \bibinfo {author}
  {\bibfnamefont {J.}~\bibnamefont {Morrison}}, \bibinfo {author}
  {\bibfnamefont {R.}~\bibnamefont {Nedbailo}}, \bibinfo {author}
  {\bibfnamefont {A.}~\bibnamefont {Picksley}}, \bibinfo {author}
  {\bibfnamefont {H.}~\bibnamefont {Song}}, \bibinfo {author} {\bibfnamefont
  {S.}~\bibnamefont {Wang}}, \bibinfo {author} {\bibfnamefont {J.~J.}\
  \bibnamefont {Rocca}},\ and\ \bibinfo {author} {\bibfnamefont {H.~M.}\
  \bibnamefont {Milchberg}},\ }\bibfield  {title} {\enquote {\bibinfo {title}
  {Multi-gev electron bunches from an all-optical laser wakefield
  accelerator},}\ }\href {https://doi.org/10.1103/PhysRevX.12.031038}
  {\bibfield  {journal} {\bibinfo  {journal} {Phys. Rev. X}\ }\textbf {\bibinfo
  {volume} {12}},\ \bibinfo {pages} {031038} (\bibinfo {year}
  {2022})}\BibitemShut {NoStop}%
\bibitem [{\citenamefont {Picksley}\ \emph {et~al.}(2024)\citenamefont
  {Picksley}, \citenamefont {Stackhouse}, \citenamefont {Benedetti},
  \citenamefont {Nakamura}, \citenamefont {Tsai}, \citenamefont {Li},
  \citenamefont {Miao}, \citenamefont {Shrock}, \citenamefont {Rockafellow},
  \citenamefont {Milchberg}, \citenamefont {Schroeder}, \citenamefont {van
  Tilborg}, \citenamefont {Esarey}, \citenamefont {Geddes},\ and\ \citenamefont
  {Gonsalves}}]{PhysRevLett.133.255001}%
  \BibitemOpen
  \bibfield  {author} {\bibinfo {author} {\bibfnamefont {A.}~\bibnamefont
  {Picksley}}, \bibinfo {author} {\bibfnamefont {J.}~\bibnamefont
  {Stackhouse}}, \bibinfo {author} {\bibfnamefont {C.}~\bibnamefont
  {Benedetti}}, \bibinfo {author} {\bibfnamefont {K.}~\bibnamefont {Nakamura}},
  \bibinfo {author} {\bibfnamefont {H.~E.}\ \bibnamefont {Tsai}}, \bibinfo
  {author} {\bibfnamefont {R.}~\bibnamefont {Li}}, \bibinfo {author}
  {\bibfnamefont {B.}~\bibnamefont {Miao}}, \bibinfo {author} {\bibfnamefont
  {J.~E.}\ \bibnamefont {Shrock}}, \bibinfo {author} {\bibfnamefont
  {E.}~\bibnamefont {Rockafellow}}, \bibinfo {author} {\bibfnamefont {H.~M.}\
  \bibnamefont {Milchberg}}, \bibinfo {author} {\bibfnamefont {C.~B.}\
  \bibnamefont {Schroeder}}, \bibinfo {author} {\bibfnamefont {J.}~\bibnamefont
  {van Tilborg}}, \bibinfo {author} {\bibfnamefont {E.}~\bibnamefont {Esarey}},
  \bibinfo {author} {\bibfnamefont {C.~G.~R.}\ \bibnamefont {Geddes}},\ and\
  \bibinfo {author} {\bibfnamefont {A.~J.}\ \bibnamefont {Gonsalves}},\
  }\bibfield  {title} {\enquote {\bibinfo {title} {Matched guiding and
  controlled injection in dark-current-free, 10-gev-class, channel-guided
  laser-plasma accelerators},}\ }\href
  {https://doi.org/10.1103/PhysRevLett.133.255001} {\bibfield  {journal}
  {\bibinfo  {journal} {Phys. Rev. Lett.}\ }\textbf {\bibinfo {volume} {133}},\
  \bibinfo {pages} {255001} (\bibinfo {year} {2024})}\BibitemShut {NoStop}%
\bibitem [{\citenamefont {Nicks}\ \emph {et~al.}()\citenamefont {Nicks},
  \citenamefont {Tajima}, \citenamefont {Roa}, \citenamefont {Nečas},\ and\
  \citenamefont {Mourou}}]{Nicks}%
  \BibitemOpen
  \bibfield  {author} {\bibinfo {author} {\bibfnamefont {B.~S.}\ \bibnamefont
  {Nicks}}, \bibinfo {author} {\bibfnamefont {T.}~\bibnamefont {Tajima}},
  \bibinfo {author} {\bibfnamefont {D.}~\bibnamefont {Roa}}, \bibinfo {author}
  {\bibfnamefont {A.}~\bibnamefont {Nečas}},\ and\ \bibinfo {author}
  {\bibfnamefont {G.}~\bibnamefont {Mourou}},\ }\enquote {\bibinfo {title}
  {Laser-wakefield application to oncology},}\ in\ \href
  {https://doi.org/10.1142/9789811217135_0016} {\emph {\bibinfo {booktitle}
  {Beam Acceleration in Crystals and Nanostructures}}},\ pp.\ \bibinfo {pages}
  {223--236}\BibitemShut {NoStop}%
\bibitem [{\citenamefont {Roa}\ \emph {et~al.}(2022)\citenamefont {Roa},
  \citenamefont {Kuo}, \citenamefont {Moyses}, \citenamefont {Taborek},
  \citenamefont {Tajima}, \citenamefont {Mourou},\ and\ \citenamefont
  {Tamanoi}}]{Roa}%
  \BibitemOpen
  \bibfield  {author} {\bibinfo {author} {\bibfnamefont {D.}~\bibnamefont
  {Roa}}, \bibinfo {author} {\bibfnamefont {J.}~\bibnamefont {Kuo}}, \bibinfo
  {author} {\bibfnamefont {H.}~\bibnamefont {Moyses}}, \bibinfo {author}
  {\bibfnamefont {P.}~\bibnamefont {Taborek}}, \bibinfo {author} {\bibfnamefont
  {T.}~\bibnamefont {Tajima}}, \bibinfo {author} {\bibfnamefont
  {G.}~\bibnamefont {Mourou}},\ and\ \bibinfo {author} {\bibfnamefont
  {F.}~\bibnamefont {Tamanoi}},\ }\bibfield  {title} {\enquote {\bibinfo
  {title} {Fiber-optic based laser wakefield accelerated electron beams and
  potential applications in radiotherapy cancer treatments},}\ }\href
  {https://doi.org/10.3390/photonics9060403} {\bibfield  {journal} {\bibinfo
  {journal} {Photonics}\ }\textbf {\bibinfo {volume} {9}} (\bibinfo {year}
  {2022}),\ 10.3390/photonics9060403}\BibitemShut {NoStop}%
\bibitem [{\citenamefont {Gahn}\ \emph {et~al.}(1999)\citenamefont {Gahn},
  \citenamefont {Tsakiris}, \citenamefont {Pukhov}, \citenamefont {Meyer-ter
  Vehn}, \citenamefont {Pretzler}, \citenamefont {Thirolf}, \citenamefont
  {Habs},\ and\ \citenamefont {Witte}}]{PhysRevLett.83.4772}%
  \BibitemOpen
  \bibfield  {author} {\bibinfo {author} {\bibfnamefont {C.}~\bibnamefont
  {Gahn}}, \bibinfo {author} {\bibfnamefont {G.~D.}\ \bibnamefont {Tsakiris}},
  \bibinfo {author} {\bibfnamefont {A.}~\bibnamefont {Pukhov}}, \bibinfo
  {author} {\bibfnamefont {J.}~\bibnamefont {Meyer-ter Vehn}}, \bibinfo
  {author} {\bibfnamefont {G.}~\bibnamefont {Pretzler}}, \bibinfo {author}
  {\bibfnamefont {P.}~\bibnamefont {Thirolf}}, \bibinfo {author} {\bibfnamefont
  {D.}~\bibnamefont {Habs}},\ and\ \bibinfo {author} {\bibfnamefont {K.~J.}\
  \bibnamefont {Witte}},\ }\bibfield  {title} {\enquote {\bibinfo {title}
  {Multi-mev electron beam generation by direct laser acceleration in
  high-density plasma channels},}\ }\href
  {https://doi.org/10.1103/PhysRevLett.83.4772} {\bibfield  {journal} {\bibinfo
   {journal} {Phys. Rev. Lett.}\ }\textbf {\bibinfo {volume} {83}},\ \bibinfo
  {pages} {4772--4775} (\bibinfo {year} {1999})}\BibitemShut {NoStop}%
\bibitem [{\citenamefont {Malka}\ \emph {et~al.}(1997)\citenamefont {Malka},
  \citenamefont {Fuchs}, \citenamefont {Amiranoff}, \citenamefont {Baton},
  \citenamefont {Gaillard}, \citenamefont {Miquel}, \citenamefont {P\'epin},
  \citenamefont {Rousseaux}, \citenamefont {Bonnaud}, \citenamefont {Busquet},\
  and\ \citenamefont {Lours}}]{PhysRevLett.79.2053}%
  \BibitemOpen
  \bibfield  {author} {\bibinfo {author} {\bibfnamefont {G.}~\bibnamefont
  {Malka}}, \bibinfo {author} {\bibfnamefont {J.}~\bibnamefont {Fuchs}},
  \bibinfo {author} {\bibfnamefont {F.}~\bibnamefont {Amiranoff}}, \bibinfo
  {author} {\bibfnamefont {S.~D.}\ \bibnamefont {Baton}}, \bibinfo {author}
  {\bibfnamefont {R.}~\bibnamefont {Gaillard}}, \bibinfo {author}
  {\bibfnamefont {J.~L.}\ \bibnamefont {Miquel}}, \bibinfo {author}
  {\bibfnamefont {H.}~\bibnamefont {P\'epin}}, \bibinfo {author} {\bibfnamefont
  {C.}~\bibnamefont {Rousseaux}}, \bibinfo {author} {\bibfnamefont
  {G.}~\bibnamefont {Bonnaud}}, \bibinfo {author} {\bibfnamefont
  {M.}~\bibnamefont {Busquet}},\ and\ \bibinfo {author} {\bibfnamefont
  {L.}~\bibnamefont {Lours}},\ }\bibfield  {title} {\enquote {\bibinfo {title}
  {Suprathermal electron generation and channel formation by an
  ultrarelativistic laser pulse in an underdense preformed plasma},}\ }\href
  {https://doi.org/10.1103/PhysRevLett.79.2053} {\bibfield  {journal} {\bibinfo
   {journal} {Phys. Rev. Lett.}\ }\textbf {\bibinfo {volume} {79}},\ \bibinfo
  {pages} {2053--2056} (\bibinfo {year} {1997})}\BibitemShut {NoStop}%
\bibitem [{\citenamefont {Pukhov}, \citenamefont {Sheng},\ and\ \citenamefont
  {Meyer-ter Vehn}(1999)}]{10.1063/1.873242}%
  \BibitemOpen
  \bibfield  {author} {\bibinfo {author} {\bibfnamefont {A.}~\bibnamefont
  {Pukhov}}, \bibinfo {author} {\bibfnamefont {Z.-M.}\ \bibnamefont {Sheng}},\
  and\ \bibinfo {author} {\bibfnamefont {J.}~\bibnamefont {Meyer-ter Vehn}},\
  }\bibfield  {title} {\enquote {\bibinfo {title} {Particle acceleration in
  relativistic laser channels},}\ }\href {https://doi.org/10.1063/1.873242}
  {\bibfield  {journal} {\bibinfo  {journal} {Physics of Plasmas}\ }\textbf
  {\bibinfo {volume} {6}},\ \bibinfo {pages} {2847--2854} (\bibinfo {year}
  {1999})}\BibitemShut {NoStop}%
\bibitem [{\citenamefont {Hussein}\ \emph {et~al.}(2021)\citenamefont
  {Hussein}, \citenamefont {Arefiev}, \citenamefont {Batson}, \citenamefont
  {Chen}, \citenamefont {Craxton}, \citenamefont {Davies}, \citenamefont
  {Froula}, \citenamefont {Gong}, \citenamefont {Haberberger}, \citenamefont
  {Ma}, \citenamefont {Nilson}, \citenamefont {Theobald}, \citenamefont {Wang},
  \citenamefont {Weichman}, \citenamefont {Williams},\ and\ \citenamefont
  {Willingale}}]{Hussein_2021}%
  \BibitemOpen
  \bibfield  {author} {\bibinfo {author} {\bibfnamefont {A.~E.}\ \bibnamefont
  {Hussein}}, \bibinfo {author} {\bibfnamefont {A.~V.}\ \bibnamefont
  {Arefiev}}, \bibinfo {author} {\bibfnamefont {T.}~\bibnamefont {Batson}},
  \bibinfo {author} {\bibfnamefont {H.}~\bibnamefont {Chen}}, \bibinfo {author}
  {\bibfnamefont {R.~S.}\ \bibnamefont {Craxton}}, \bibinfo {author}
  {\bibfnamefont {A.~S.}\ \bibnamefont {Davies}}, \bibinfo {author}
  {\bibfnamefont {D.~H.}\ \bibnamefont {Froula}}, \bibinfo {author}
  {\bibfnamefont {Z.}~\bibnamefont {Gong}}, \bibinfo {author} {\bibfnamefont
  {D.}~\bibnamefont {Haberberger}}, \bibinfo {author} {\bibfnamefont
  {Y.}~\bibnamefont {Ma}}, \bibinfo {author} {\bibfnamefont {P.~M.}\
  \bibnamefont {Nilson}}, \bibinfo {author} {\bibfnamefont {W.}~\bibnamefont
  {Theobald}}, \bibinfo {author} {\bibfnamefont {T.}~\bibnamefont {Wang}},
  \bibinfo {author} {\bibfnamefont {K.}~\bibnamefont {Weichman}}, \bibinfo
  {author} {\bibfnamefont {G.~J.}\ \bibnamefont {Williams}},\ and\ \bibinfo
  {author} {\bibfnamefont {L.}~\bibnamefont {Willingale}},\ }\bibfield  {title}
  {\enquote {\bibinfo {title} {Towards the optimisation of direct laser
  acceleration},}\ }\href {https://doi.org/10.1088/1367-2630/abdf9a} {\bibfield
   {journal} {\bibinfo  {journal} {New Journal of Physics}\ }\textbf {\bibinfo
  {volume} {23}},\ \bibinfo {pages} {023031} (\bibinfo {year}
  {2021})}\BibitemShut {NoStop}%
\bibitem [{\citenamefont {Cohen}\ \emph {et~al.}(2024)\citenamefont {Cohen},
  \citenamefont {Meir}, \citenamefont {Tangtartharakul}, \citenamefont
  {Perelmutter}, \citenamefont {Elkind}, \citenamefont {Gershuni},
  \citenamefont {Levanon}, \citenamefont {Arefiev},\ and\ \citenamefont
  {Pomerantz}}]{doi:10.1126/sciadv.adk1947}%
  \BibitemOpen
  \bibfield  {author} {\bibinfo {author} {\bibfnamefont {I.}~\bibnamefont
  {Cohen}}, \bibinfo {author} {\bibfnamefont {T.}~\bibnamefont {Meir}},
  \bibinfo {author} {\bibfnamefont {K.}~\bibnamefont {Tangtartharakul}},
  \bibinfo {author} {\bibfnamefont {L.}~\bibnamefont {Perelmutter}}, \bibinfo
  {author} {\bibfnamefont {M.}~\bibnamefont {Elkind}}, \bibinfo {author}
  {\bibfnamefont {Y.}~\bibnamefont {Gershuni}}, \bibinfo {author}
  {\bibfnamefont {A.}~\bibnamefont {Levanon}}, \bibinfo {author} {\bibfnamefont
  {A.~V.}\ \bibnamefont {Arefiev}},\ and\ \bibinfo {author} {\bibfnamefont
  {I.}~\bibnamefont {Pomerantz}},\ }\bibfield  {title} {\enquote {\bibinfo
  {title} {Undepleted direct laser acceleration},}\ }\href
  {https://doi.org/10.1126/sciadv.adk1947} {\bibfield  {journal} {\bibinfo
  {journal} {Science Advances}\ }\textbf {\bibinfo {volume} {10}},\ \bibinfo
  {pages} {eadk1947} (\bibinfo {year} {2024})}\BibitemShut {NoStop}%
\bibitem [{\citenamefont {Salamin}\ and\ \citenamefont
  {Keitel}(2002)}]{PhysRevLett.88.095005}%
  \BibitemOpen
  \bibfield  {author} {\bibinfo {author} {\bibfnamefont {Y.~I.}\ \bibnamefont
  {Salamin}}\ and\ \bibinfo {author} {\bibfnamefont {C.~H.}\ \bibnamefont
  {Keitel}},\ }\bibfield  {title} {\enquote {\bibinfo {title} {Electron
  acceleration by a tightly focused laser beam},}\ }\href
  {https://doi.org/10.1103/PhysRevLett.88.095005} {\bibfield  {journal}
  {\bibinfo  {journal} {Phys. Rev. Lett.}\ }\textbf {\bibinfo {volume} {88}},\
  \bibinfo {pages} {095005} (\bibinfo {year} {2002})}\BibitemShut {NoStop}%
\bibitem [{\citenamefont {He}\ \emph {et~al.}(2003)\citenamefont {He},
  \citenamefont {Yu}, \citenamefont {Lu}, \citenamefont {Xu}, \citenamefont
  {Qian}, \citenamefont {Shen}, \citenamefont {Yuan}, \citenamefont {Li},\ and\
  \citenamefont {Xu}}]{PhysRevE.68.046407}%
  \BibitemOpen
  \bibfield  {author} {\bibinfo {author} {\bibfnamefont {F.}~\bibnamefont
  {He}}, \bibinfo {author} {\bibfnamefont {W.}~\bibnamefont {Yu}}, \bibinfo
  {author} {\bibfnamefont {P.}~\bibnamefont {Lu}}, \bibinfo {author}
  {\bibfnamefont {H.}~\bibnamefont {Xu}}, \bibinfo {author} {\bibfnamefont
  {L.}~\bibnamefont {Qian}}, \bibinfo {author} {\bibfnamefont {B.}~\bibnamefont
  {Shen}}, \bibinfo {author} {\bibfnamefont {X.}~\bibnamefont {Yuan}}, \bibinfo
  {author} {\bibfnamefont {R.}~\bibnamefont {Li}},\ and\ \bibinfo {author}
  {\bibfnamefont {Z.}~\bibnamefont {Xu}},\ }\bibfield  {title} {\enquote
  {\bibinfo {title} {Ponderomotive acceleration of electrons by a tightly
  focused intense laser beam},}\ }\href
  {https://doi.org/10.1103/PhysRevE.68.046407} {\bibfield  {journal} {\bibinfo
  {journal} {Phys. Rev. E}\ }\textbf {\bibinfo {volume} {68}},\ \bibinfo
  {pages} {046407} (\bibinfo {year} {2003})}\BibitemShut {NoStop}%
\bibitem [{\citenamefont {Popov}\ \emph {et~al.}(2008)\citenamefont {Popov},
  \citenamefont {Bychenkov}, \citenamefont {Rozmus},\ and\ \citenamefont
  {Sydora}}]{10.1063/1.2830651}%
  \BibitemOpen
  \bibfield  {author} {\bibinfo {author} {\bibfnamefont {K.~I.}\ \bibnamefont
  {Popov}}, \bibinfo {author} {\bibfnamefont {V.~Y.}\ \bibnamefont
  {Bychenkov}}, \bibinfo {author} {\bibfnamefont {W.}~\bibnamefont {Rozmus}},\
  and\ \bibinfo {author} {\bibfnamefont {R.~D.}\ \bibnamefont {Sydora}},\
  }\bibfield  {title} {\enquote {\bibinfo {title} {Electron vacuum acceleration
  by a tightly focused laser pulse},}\ }\href
  {https://doi.org/10.1063/1.2830651} {\bibfield  {journal} {\bibinfo
  {journal} {Physics of Plasmas}\ }\textbf {\bibinfo {volume} {15}},\ \bibinfo
  {pages} {013108} (\bibinfo {year} {2008})}\BibitemShut {NoStop}%
\bibitem [{\citenamefont {Varin}\ and\ \citenamefont
  {Pich{\'e}}(2002)}]{varin2002acceleration}%
  \BibitemOpen
  \bibfield  {author} {\bibinfo {author} {\bibfnamefont {C.}~\bibnamefont
  {Varin}}\ and\ \bibinfo {author} {\bibfnamefont {M.}~\bibnamefont
  {Pich{\'e}}},\ }\bibfield  {title} {\enquote {\bibinfo {title} {Acceleration
  of ultra-relativistic electrons using high-intensity tm01 laser beams},}\
  }\href {https://doi.org/https://doi.org/10.1007/s00340-002-0906-8} {\bibfield
   {journal} {\bibinfo  {journal} {Applied physics B}\ }\textbf {\bibinfo
  {volume} {74}},\ \bibinfo {pages} {s83--s88} (\bibinfo {year}
  {2002})}\BibitemShut {NoStop}%
\bibitem [{\citenamefont {Bochkarev}\ and\ \citenamefont
  {Bychenkov}(2007)}]{Bochkarev_2007}%
  \BibitemOpen
  \bibfield  {author} {\bibinfo {author} {\bibfnamefont {S.~G.}\ \bibnamefont
  {Bochkarev}}\ and\ \bibinfo {author} {\bibfnamefont {V.~Y.}\ \bibnamefont
  {Bychenkov}},\ }\bibfield  {title} {\enquote {\bibinfo {title} {Acceleration
  of electrons by tightly focused femtosecond laser pulses},}\ }\href
  {https://doi.org/10.1070/QE2007v037n03ABEH013462} {\bibfield  {journal}
  {\bibinfo  {journal} {Quantum Electronics}\ }\textbf {\bibinfo {volume}
  {37}},\ \bibinfo {pages} {273} (\bibinfo {year} {2007})}\BibitemShut
  {NoStop}%
\bibitem [{\citenamefont {Varin}\ \emph {et~al.}(2013)\citenamefont {Varin},
  \citenamefont {Payeur}, \citenamefont {Marceau}, \citenamefont {Fourmaux},
  \citenamefont {April}, \citenamefont {Schmidt}, \citenamefont {Fortin},
  \citenamefont {Thiré}, \citenamefont {Brabec}, \citenamefont {Légaré},
  \citenamefont {Kieffer},\ and\ \citenamefont {Piché}}]{app3010070}%
  \BibitemOpen
  \bibfield  {author} {\bibinfo {author} {\bibfnamefont {C.}~\bibnamefont
  {Varin}}, \bibinfo {author} {\bibfnamefont {S.}~\bibnamefont {Payeur}},
  \bibinfo {author} {\bibfnamefont {V.}~\bibnamefont {Marceau}}, \bibinfo
  {author} {\bibfnamefont {S.}~\bibnamefont {Fourmaux}}, \bibinfo {author}
  {\bibfnamefont {A.}~\bibnamefont {April}}, \bibinfo {author} {\bibfnamefont
  {B.}~\bibnamefont {Schmidt}}, \bibinfo {author} {\bibfnamefont {P.-L.}\
  \bibnamefont {Fortin}}, \bibinfo {author} {\bibfnamefont {N.}~\bibnamefont
  {Thiré}}, \bibinfo {author} {\bibfnamefont {T.}~\bibnamefont {Brabec}},
  \bibinfo {author} {\bibfnamefont {F.}~\bibnamefont {Légaré}}, \bibinfo
  {author} {\bibfnamefont {J.-C.}\ \bibnamefont {Kieffer}},\ and\ \bibinfo
  {author} {\bibfnamefont {M.}~\bibnamefont {Piché}},\ }\bibfield  {title}
  {\enquote {\bibinfo {title} {Direct electron acceleration with radially
  polarized laser beams},}\ }\href {https://doi.org/10.3390/app3010070}
  {\bibfield  {journal} {\bibinfo  {journal} {Applied Sciences}\ }\textbf
  {\bibinfo {volume} {3}},\ \bibinfo {pages} {70--93} (\bibinfo {year}
  {2013})}\BibitemShut {NoStop}%
\bibitem [{\citenamefont {Marceau}\ \emph {et~al.}(2013)\citenamefont
  {Marceau}, \citenamefont {Varin}, \citenamefont {Brabec},\ and\ \citenamefont
  {Pich\'e}}]{PhysRevLett.111.224801}%
  \BibitemOpen
  \bibfield  {author} {\bibinfo {author} {\bibfnamefont {V.}~\bibnamefont
  {Marceau}}, \bibinfo {author} {\bibfnamefont {C.}~\bibnamefont {Varin}},
  \bibinfo {author} {\bibfnamefont {T.}~\bibnamefont {Brabec}},\ and\ \bibinfo
  {author} {\bibfnamefont {M.}~\bibnamefont {Pich\'e}},\ }\bibfield  {title}
  {\enquote {\bibinfo {title} {Femtosecond 240-kev electron pulses from direct
  laser acceleration in a low-density gas},}\ }\href
  {https://doi.org/10.1103/PhysRevLett.111.224801} {\bibfield  {journal}
  {\bibinfo  {journal} {Phys. Rev. Lett.}\ }\textbf {\bibinfo {volume} {111}},\
  \bibinfo {pages} {224801} (\bibinfo {year} {2013})}\BibitemShut {NoStop}%
\bibitem [{\citenamefont {Payeur}\ \emph {et~al.}(2012)\citenamefont {Payeur},
  \citenamefont {Fourmaux}, \citenamefont {Schmidt}, \citenamefont {MacLean},
  \citenamefont {Tchervenkov}, \citenamefont {Légaré}, \citenamefont
  {Piché},\ and\ \citenamefont {Kieffer}}]{Payeur}%
  \BibitemOpen
  \bibfield  {author} {\bibinfo {author} {\bibfnamefont {S.}~\bibnamefont
  {Payeur}}, \bibinfo {author} {\bibfnamefont {S.}~\bibnamefont {Fourmaux}},
  \bibinfo {author} {\bibfnamefont {B.~E.}\ \bibnamefont {Schmidt}}, \bibinfo
  {author} {\bibfnamefont {J.~P.}\ \bibnamefont {MacLean}}, \bibinfo {author}
  {\bibfnamefont {C.}~\bibnamefont {Tchervenkov}}, \bibinfo {author}
  {\bibfnamefont {F.}~\bibnamefont {Légaré}}, \bibinfo {author}
  {\bibfnamefont {M.}~\bibnamefont {Piché}},\ and\ \bibinfo {author}
  {\bibfnamefont {J.~C.}\ \bibnamefont {Kieffer}},\ }\bibfield  {title}
  {\enquote {\bibinfo {title} {Generation of a beam of fast electrons by
  tightly focusing a radially polarized ultrashort laser pulse},}\ }\href
  {https://doi.org/10.1063/1.4738998} {\bibfield  {journal} {\bibinfo
  {journal} {Applied Physics Letters}\ }\textbf {\bibinfo {volume} {101}},\
  \bibinfo {pages} {041105} (\bibinfo {year} {2012})},\ \Eprint
  {https://arxiv.org/abs/https://pubs.aip.org/aip/apl/article-pdf/doi/10.1063/1.4738998/14260845/041105\_1\_online.pdf}
  {https://pubs.aip.org/aip/apl/article-pdf/doi/10.1063/1.4738998/14260845/041105\_1\_online.pdf}
  \BibitemShut {NoStop}%
\bibitem [{\citenamefont {Carbajo}\ \emph {et~al.}(2016)\citenamefont
  {Carbajo}, \citenamefont {Nanni}, \citenamefont {Wong}, \citenamefont
  {Moriena}, \citenamefont {Keathley}, \citenamefont {Laurent}, \citenamefont
  {Miller},\ and\ \citenamefont {K\"artner}}]{PhysRevAccelBeams.19.021303}%
  \BibitemOpen
  \bibfield  {author} {\bibinfo {author} {\bibfnamefont {S.}~\bibnamefont
  {Carbajo}}, \bibinfo {author} {\bibfnamefont {E.~A.}\ \bibnamefont {Nanni}},
  \bibinfo {author} {\bibfnamefont {L.~J.}\ \bibnamefont {Wong}}, \bibinfo
  {author} {\bibfnamefont {G.}~\bibnamefont {Moriena}}, \bibinfo {author}
  {\bibfnamefont {P.~D.}\ \bibnamefont {Keathley}}, \bibinfo {author}
  {\bibfnamefont {G.}~\bibnamefont {Laurent}}, \bibinfo {author} {\bibfnamefont
  {R.~J.~D.}\ \bibnamefont {Miller}},\ and\ \bibinfo {author} {\bibfnamefont
  {F.~X.}\ \bibnamefont {K\"artner}},\ }\bibfield  {title} {\enquote {\bibinfo
  {title} {Direct longitudinal laser acceleration of electrons in free
  space},}\ }\href {https://doi.org/10.1103/PhysRevAccelBeams.19.021303}
  {\bibfield  {journal} {\bibinfo  {journal} {Phys. Rev. Accel. Beams}\
  }\textbf {\bibinfo {volume} {19}},\ \bibinfo {pages} {021303} (\bibinfo
  {year} {2016})}\BibitemShut {NoStop}%
\bibitem [{\citenamefont {Powell}\ \emph {et~al.}(2024)\citenamefont {Powell},
  \citenamefont {Jolly}, \citenamefont {Valli\`eres}, \citenamefont
  {Fillion-Gourdeau}, \citenamefont {Payeur}, \citenamefont {Fourmaux},
  \citenamefont {Lytova}, \citenamefont {Pich\'e}, \citenamefont {Ibrahim},
  \citenamefont {MacLean},\ and\ \citenamefont
  {L\'egar\'e}}]{powell2024relativistic}%
  \BibitemOpen
  \bibfield  {author} {\bibinfo {author} {\bibfnamefont {J.}~\bibnamefont
  {Powell}}, \bibinfo {author} {\bibfnamefont {S.~W.}\ \bibnamefont {Jolly}},
  \bibinfo {author} {\bibfnamefont {S.}~\bibnamefont {Valli\`eres}}, \bibinfo
  {author} {\bibfnamefont {F.~m.~c.}\ \bibnamefont {Fillion-Gourdeau}},
  \bibinfo {author} {\bibfnamefont {S.}~\bibnamefont {Payeur}}, \bibinfo
  {author} {\bibfnamefont {S.}~\bibnamefont {Fourmaux}}, \bibinfo {author}
  {\bibfnamefont {M.}~\bibnamefont {Lytova}}, \bibinfo {author} {\bibfnamefont
  {M.}~\bibnamefont {Pich\'e}}, \bibinfo {author} {\bibfnamefont
  {H.}~\bibnamefont {Ibrahim}}, \bibinfo {author} {\bibfnamefont
  {S.}~\bibnamefont {MacLean}},\ and\ \bibinfo {author} {\bibfnamefont
  {F.~m.~c.}\ \bibnamefont {L\'egar\'e}},\ }\bibfield  {title} {\enquote
  {\bibinfo {title} {Relativistic electrons from vacuum laser acceleration
  using tightly focused radially polarized beams},}\ }\href
  {https://doi.org/10.1103/PhysRevLett.133.155001} {\bibfield  {journal}
  {\bibinfo  {journal} {Phys. Rev. Lett.}\ }\textbf {\bibinfo {volume} {133}},\
  \bibinfo {pages} {155001} (\bibinfo {year} {2024})}\BibitemShut {NoStop}%
\bibitem [{\citenamefont {Valli\`{e}res}\ \emph {et~al.}(2023)\citenamefont
  {Valli\`{e}res}, \citenamefont {Fillion-Gourdeau}, \citenamefont {Payeur},
  \citenamefont {Powell}, \citenamefont {Fourmaux}, \citenamefont
  {L\'{e}gar\'{e}},\ and\ \citenamefont {Maclean}}]{Vallieres:23}%
  \BibitemOpen
  \bibfield  {author} {\bibinfo {author} {\bibfnamefont {S.}~\bibnamefont
  {Valli\`{e}res}}, \bibinfo {author} {\bibfnamefont {F.}~\bibnamefont
  {Fillion-Gourdeau}}, \bibinfo {author} {\bibfnamefont {S.}~\bibnamefont
  {Payeur}}, \bibinfo {author} {\bibfnamefont {J.}~\bibnamefont {Powell}},
  \bibinfo {author} {\bibfnamefont {S.}~\bibnamefont {Fourmaux}}, \bibinfo
  {author} {\bibfnamefont {F.}~\bibnamefont {L\'{e}gar\'{e}}},\ and\ \bibinfo
  {author} {\bibfnamefont {S.}~\bibnamefont {Maclean}},\ }\bibfield  {title}
  {\enquote {\bibinfo {title} {Tight-focusing parabolic reflector schemes for
  petawatt lasers},}\ }\href {https://doi.org/10.1364/OE.486230} {\bibfield
  {journal} {\bibinfo  {journal} {Opt. Express}\ }\textbf {\bibinfo {volume}
  {31}},\ \bibinfo {pages} {19319--19335} (\bibinfo {year} {2023})}\BibitemShut
  {NoStop}%
\bibitem [{\citenamefont {Landau}\ and\ \citenamefont
  {Lifshitz}(1980)}]{Landau}%
  \BibitemOpen
  \bibfield  {author} {\bibinfo {author} {\bibfnamefont {L.}~\bibnamefont
  {Landau}}\ and\ \bibinfo {author} {\bibfnamefont {E.}~\bibnamefont
  {Lifshitz}},\ }\href {http://www.worldcat.org/isbn/0750627689} {\emph
  {\bibinfo {title} {The Classical Theory of Fields}}},\ \bibinfo {edition}
  {4th}\ ed.,\ Course of theoretical physics: Volume 2\ (\bibinfo  {publisher}
  {Butterworth-Heinemann},\ \bibinfo {year} {1980})\BibitemShut {NoStop}%
\bibitem [{\citenamefont {Stupakov}\ and\ \citenamefont
  {Zolotorev}(2001)}]{Stupakov}%
  \BibitemOpen
  \bibfield  {author} {\bibinfo {author} {\bibfnamefont {G.~V.}\ \bibnamefont
  {Stupakov}}\ and\ \bibinfo {author} {\bibfnamefont {M.~S.}\ \bibnamefont
  {Zolotorev}},\ }\bibfield  {title} {\enquote {\bibinfo {title} {Ponderomotive
  laser acceleration and focusing in vacuum for generation of attosecond
  electron bunches},}\ }\href {https://doi.org/10.1103/PhysRevLett.86.5274}
  {\bibfield  {journal} {\bibinfo  {journal} {Phys. Rev. Lett.}\ }\textbf
  {\bibinfo {volume} {86}},\ \bibinfo {pages} {5274--5277} (\bibinfo {year}
  {2001})}\BibitemShut {NoStop}%
\bibitem [{\citenamefont {Bauer}, \citenamefont {Mulser},\ and\ \citenamefont
  {Steeb}(1995)}]{Bauer}%
  \BibitemOpen
  \bibfield  {author} {\bibinfo {author} {\bibfnamefont {D.}~\bibnamefont
  {Bauer}}, \bibinfo {author} {\bibfnamefont {P.}~\bibnamefont {Mulser}},\ and\
  \bibinfo {author} {\bibfnamefont {W.~H.}\ \bibnamefont {Steeb}},\ }\bibfield
  {title} {\enquote {\bibinfo {title} {Relativistic ponderomotive force, uphill
  acceleration, and transition to chaos},}\ }\href
  {https://doi.org/10.1103/PhysRevLett.75.4622} {\bibfield  {journal} {\bibinfo
   {journal} {Phys. Rev. Lett.}\ }\textbf {\bibinfo {volume} {75}},\ \bibinfo
  {pages} {4622--4625} (\bibinfo {year} {1995})}\BibitemShut {NoStop}%
\bibitem [{\citenamefont {April}(2010)}]{April_2010}%
  \BibitemOpen
  \bibfield  {author} {\bibinfo {author} {\bibfnamefont {A.}~\bibnamefont
  {April}},\ }\enquote {\bibinfo {title} {Coherence and ultrashort pulse laser
  emission},}\ \ (\bibinfo  {publisher} {IntechOpen},\ \bibinfo {year} {2010})\
  pp.\ \bibinfo {pages} {355–--382}\BibitemShut {NoStop}%
\bibitem [{\citenamefont {Jolly}\ \emph {et~al.}(2025)\citenamefont {Jolly},
  \citenamefont {Lytova}, \citenamefont {Vallières}, \citenamefont {Légaré},
  \citenamefont {MacLean},\ and\ \citenamefont {Fillion-Gourdeau}}]{Jolly}%
  \BibitemOpen
  \bibfield  {author} {\bibinfo {author} {\bibfnamefont {S.~W.}\ \bibnamefont
  {Jolly}}, \bibinfo {author} {\bibfnamefont {M.}~\bibnamefont {Lytova}},
  \bibinfo {author} {\bibfnamefont {S.}~\bibnamefont {Vallières}}, \bibinfo
  {author} {\bibfnamefont {F.}~\bibnamefont {Légaré}}, \bibinfo {author}
  {\bibfnamefont {S.}~\bibnamefont {MacLean}},\ and\ \bibinfo {author}
  {\bibfnamefont {F.}~\bibnamefont {Fillion-Gourdeau}},\ }\bibfield  {title}
  {\enquote {\bibinfo {title} {Space-time couplings in ultrashort lasers with
  arbitrary nonparaxial focusing},}\ }\href
  {https://doi.org/doi:10.1515/nanoph-2024-0616} {\bibfield  {journal}
  {\bibinfo  {journal} {Nanophotonics}\ }\textbf {\bibinfo {volume} {14}},\
  \bibinfo {pages} {815--832} (\bibinfo {year} {2025})}\BibitemShut {NoStop}%
\bibitem [{\citenamefont {Dumont}\ \emph {et~al.}(2017)\citenamefont {Dumont},
  \citenamefont {Fillion-Gourdeau}, \citenamefont {Lefebvre}, \citenamefont
  {Gagnon},\ and\ \citenamefont {MacLean}}]{Dumont_2017}%
  \BibitemOpen
  \bibfield  {author} {\bibinfo {author} {\bibfnamefont {J.}~\bibnamefont
  {Dumont}}, \bibinfo {author} {\bibfnamefont {F.}~\bibnamefont
  {Fillion-Gourdeau}}, \bibinfo {author} {\bibfnamefont {C.}~\bibnamefont
  {Lefebvre}}, \bibinfo {author} {\bibfnamefont {D.}~\bibnamefont {Gagnon}},\
  and\ \bibinfo {author} {\bibfnamefont {S.}~\bibnamefont {MacLean}},\
  }\bibfield  {title} {\enquote {\bibinfo {title} {Efficiently parallelized
  modeling of tightly focused, large bandwidth laser pulses},}\ }\href
  {https://doi.org/10.1088/2040-8986/aa52e9} {\bibfield  {journal} {\bibinfo
  {journal} {Journal of Optics}\ }\textbf {\bibinfo {volume} {19}},\ \bibinfo
  {pages} {025604} (\bibinfo {year} {2017})}\BibitemShut {NoStop}%
\bibitem [{\citenamefont {Corkum}, \citenamefont {Burnett},\ and\ \citenamefont
  {Brunel}(1989)}]{Corkum}%
  \BibitemOpen
  \bibfield  {author} {\bibinfo {author} {\bibfnamefont {P.~B.}\ \bibnamefont
  {Corkum}}, \bibinfo {author} {\bibfnamefont {N.~H.}\ \bibnamefont
  {Burnett}},\ and\ \bibinfo {author} {\bibfnamefont {F.}~\bibnamefont
  {Brunel}},\ }\bibfield  {title} {\enquote {\bibinfo {title} {Above-threshold
  ionization in the long-wavelength limit},}\ }\href
  {https://doi.org/10.1103/PhysRevLett.62.1259} {\bibfield  {journal} {\bibinfo
   {journal} {Phys. Rev. Lett.}\ }\textbf {\bibinfo {volume} {62}},\ \bibinfo
  {pages} {1259--1262} (\bibinfo {year} {1989})}\BibitemShut {NoStop}%
\bibitem [{\citenamefont {Derouillat}\ \emph {et~al.}(2018)\citenamefont
  {Derouillat}, \citenamefont {Beck}, \citenamefont {Pérez}, \citenamefont
  {Vinci}, \citenamefont {Chiaramello}, \citenamefont {Grassi}, \citenamefont
  {Flé}, \citenamefont {Bouchard}, \citenamefont {Plotnikov}, \citenamefont
  {Aunai}, \citenamefont {Dargent}, \citenamefont {Riconda},\ and\
  \citenamefont {Grech}}]{Smilei18}%
  \BibitemOpen
  \bibfield  {author} {\bibinfo {author} {\bibfnamefont {J.}~\bibnamefont
  {Derouillat}}, \bibinfo {author} {\bibfnamefont {A.}~\bibnamefont {Beck}},
  \bibinfo {author} {\bibfnamefont {F.}~\bibnamefont {Pérez}}, \bibinfo
  {author} {\bibfnamefont {T.}~\bibnamefont {Vinci}}, \bibinfo {author}
  {\bibfnamefont {M.}~\bibnamefont {Chiaramello}}, \bibinfo {author}
  {\bibfnamefont {A.}~\bibnamefont {Grassi}}, \bibinfo {author} {\bibfnamefont
  {M.}~\bibnamefont {Flé}}, \bibinfo {author} {\bibfnamefont {G.}~\bibnamefont
  {Bouchard}}, \bibinfo {author} {\bibfnamefont {I.}~\bibnamefont {Plotnikov}},
  \bibinfo {author} {\bibfnamefont {N.}~\bibnamefont {Aunai}}, \bibinfo
  {author} {\bibfnamefont {J.}~\bibnamefont {Dargent}}, \bibinfo {author}
  {\bibfnamefont {C.}~\bibnamefont {Riconda}},\ and\ \bibinfo {author}
  {\bibfnamefont {M.}~\bibnamefont {Grech}},\ }\bibfield  {title} {\enquote
  {\bibinfo {title} {Smilei : A collaborative, open-source, multi-purpose
  particle-in-cell code for plasma simulation},}\ }\href
  {https://doi.org/https://doi.org/10.1016/j.cpc.2017.09.024} {\bibfield
  {journal} {\bibinfo  {journal} {Computer Physics Communications}\ }\textbf
  {\bibinfo {volume} {222}},\ \bibinfo {pages} {351--373} (\bibinfo {year}
  {2018})}\BibitemShut {NoStop}%
\bibitem [{\citenamefont {Nuter}\ \emph {et~al.}(2011)\citenamefont {Nuter},
  \citenamefont {Gremillet}, \citenamefont {Lefebvre}, \citenamefont {Lévy},
  \citenamefont {Ceccotti},\ and\ \citenamefont {Martin}}]{Nuter}%
  \BibitemOpen
  \bibfield  {author} {\bibinfo {author} {\bibfnamefont {R.}~\bibnamefont
  {Nuter}}, \bibinfo {author} {\bibfnamefont {L.}~\bibnamefont {Gremillet}},
  \bibinfo {author} {\bibfnamefont {E.}~\bibnamefont {Lefebvre}}, \bibinfo
  {author} {\bibfnamefont {A.}~\bibnamefont {Lévy}}, \bibinfo {author}
  {\bibfnamefont {T.}~\bibnamefont {Ceccotti}},\ and\ \bibinfo {author}
  {\bibfnamefont {P.}~\bibnamefont {Martin}},\ }\bibfield  {title} {\enquote
  {\bibinfo {title} {{Field ionization model implemented in Particle In Cell
  code and applied to laser-accelerated carbon ions}},}\ }\href
  {https://doi.org/10.1063/1.3559494} {\bibfield  {journal} {\bibinfo
  {journal} {Physics of Plasmas}\ }\textbf {\bibinfo {volume} {18}},\ \bibinfo
  {pages} {033107} (\bibinfo {year} {2011})}\BibitemShut {NoStop}%
\bibitem [{\citenamefont {Thiré}\ \emph {et~al.}(2015)\citenamefont {Thiré},
  \citenamefont {Beaulieu}, \citenamefont {Cardin}, \citenamefont {Laramée},
  \citenamefont {Wanie}, \citenamefont {Schmidt},\ and\ \citenamefont
  {Légaré}}]{Thire2015_10mJ5cycle}%
  \BibitemOpen
  \bibfield  {author} {\bibinfo {author} {\bibfnamefont {N.}~\bibnamefont
  {Thiré}}, \bibinfo {author} {\bibfnamefont {S.}~\bibnamefont {Beaulieu}},
  \bibinfo {author} {\bibfnamefont {V.}~\bibnamefont {Cardin}}, \bibinfo
  {author} {\bibfnamefont {A.}~\bibnamefont {Laramée}}, \bibinfo {author}
  {\bibfnamefont {V.}~\bibnamefont {Wanie}}, \bibinfo {author} {\bibfnamefont
  {B.~E.}\ \bibnamefont {Schmidt}},\ and\ \bibinfo {author} {\bibfnamefont
  {F.}~\bibnamefont {Légaré}},\ }\bibfield  {title} {\enquote {\bibinfo
  {title} {10 {mJ} 5‑cycle pulses at 1.8 {$\mu$m} through optical parametric
  amplification},}\ }\href {https://doi.org/10.1063/1.4914344} {\bibfield
  {journal} {\bibinfo  {journal} {Applied Physics Letters}\ }\textbf {\bibinfo
  {volume} {106}},\ \bibinfo {pages} {091110} (\bibinfo {year}
  {2015})}\BibitemShut {NoStop}%
\bibitem [{\citenamefont {Heuermann}\ \emph {et~al.}(2022)\citenamefont
  {Heuermann}, \citenamefont {Wang}, \citenamefont {Lenski}, \citenamefont
  {Gebhardt}, \citenamefont {Gaida}, \citenamefont {Abdelaal}, \citenamefont
  {Buldt}, \citenamefont {M\"{u}ller}, \citenamefont {Klenke},\ and\
  \citenamefont {Limpert}}]{Heuermann:22}%
  \BibitemOpen
  \bibfield  {author} {\bibinfo {author} {\bibfnamefont {T.}~\bibnamefont
  {Heuermann}}, \bibinfo {author} {\bibfnamefont {Z.}~\bibnamefont {Wang}},
  \bibinfo {author} {\bibfnamefont {M.}~\bibnamefont {Lenski}}, \bibinfo
  {author} {\bibfnamefont {M.}~\bibnamefont {Gebhardt}}, \bibinfo {author}
  {\bibfnamefont {C.}~\bibnamefont {Gaida}}, \bibinfo {author} {\bibfnamefont
  {M.}~\bibnamefont {Abdelaal}}, \bibinfo {author} {\bibfnamefont
  {J.}~\bibnamefont {Buldt}}, \bibinfo {author} {\bibfnamefont
  {M.}~\bibnamefont {M\"{u}ller}}, \bibinfo {author} {\bibfnamefont
  {A.}~\bibnamefont {Klenke}},\ and\ \bibinfo {author} {\bibfnamefont
  {J.}~\bibnamefont {Limpert}},\ }\bibfield  {title} {\enquote {\bibinfo
  {title} {Ultrafast tm-doped fiber laser system delivering 1.65-mj, sub-100-fs
  pulses at a 100-khz repetition rate},}\ }\href
  {https://doi.org/10.1364/OL.459385} {\bibfield  {journal} {\bibinfo
  {journal} {Opt. Lett.}\ }\textbf {\bibinfo {volume} {47}},\ \bibinfo {pages}
  {3095--3098} (\bibinfo {year} {2022})}\BibitemShut {NoStop}%
\bibitem [{\citenamefont {Za\"{\i}m}\ \emph {et~al.}(2020)\citenamefont
  {Za\"{\i}m}, \citenamefont {Gu\'enot}, \citenamefont {Chopineau},
  \citenamefont {Denoeud}, \citenamefont {Lundh}, \citenamefont {Vincenti},
  \citenamefont {Qu\'er\'e},\ and\ \citenamefont {Faure}}]{PhysRevX.10.041064}%
  \BibitemOpen
  \bibfield  {author} {\bibinfo {author} {\bibfnamefont {N.}~\bibnamefont
  {Za\"{\i}m}}, \bibinfo {author} {\bibfnamefont {D.}~\bibnamefont {Gu\'enot}},
  \bibinfo {author} {\bibfnamefont {L.}~\bibnamefont {Chopineau}}, \bibinfo
  {author} {\bibfnamefont {A.}~\bibnamefont {Denoeud}}, \bibinfo {author}
  {\bibfnamefont {O.}~\bibnamefont {Lundh}}, \bibinfo {author} {\bibfnamefont
  {H.}~\bibnamefont {Vincenti}}, \bibinfo {author} {\bibfnamefont
  {F.}~\bibnamefont {Qu\'er\'e}},\ and\ \bibinfo {author} {\bibfnamefont
  {J.}~\bibnamefont {Faure}},\ }\bibfield  {title} {\enquote {\bibinfo {title}
  {Interaction of ultraintense radially-polarized laser pulses with plasma
  mirrors},}\ }\href {https://doi.org/10.1103/PhysRevX.10.041064} {\bibfield
  {journal} {\bibinfo  {journal} {Phys. Rev. X}\ }\textbf {\bibinfo {volume}
  {10}},\ \bibinfo {pages} {041064} (\bibinfo {year} {2020})}\BibitemShut
  {NoStop}%
\bibitem [{\citenamefont {Jackson}(1999)}]{Jackson:100964}%
  \BibitemOpen
  \bibfield  {author} {\bibinfo {author} {\bibfnamefont {J.~D.}\ \bibnamefont
  {Jackson}},\ }\href@noop {} {\emph {\bibinfo {title} {{Classical
  electrodynamics; 3rd ed.}}}}\ (\bibinfo  {publisher} {Wiley},\ \bibinfo
  {address} {New York, NY},\ \bibinfo {year} {1999})\BibitemShut {NoStop}%
\bibitem [{\citenamefont {Lekner}(2002)}]{John_Lekner_2003}%
  \BibitemOpen
  \bibfield  {author} {\bibinfo {author} {\bibfnamefont {J.}~\bibnamefont
  {Lekner}},\ }\bibfield  {title} {\enquote {\bibinfo {title} {Polarization of
  tightly focused laser beams},}\ }\href
  {https://doi.org/10.1088/1464-4258/5/1/302} {\bibfield  {journal} {\bibinfo
  {journal} {Journal of Optics A: Pure and Applied Optics}\ }\textbf {\bibinfo
  {volume} {5}},\ \bibinfo {pages} {6} (\bibinfo {year} {2002})}\BibitemShut
  {NoStop}%
\bibitem [{\citenamefont {Caron}\ and\ \citenamefont
  {Potvliege}(1999)}]{Caron}%
  \BibitemOpen
  \bibfield  {author} {\bibinfo {author} {\bibfnamefont {C.~F.~R.}\
  \bibnamefont {Caron}}\ and\ \bibinfo {author} {\bibfnamefont {R.~M.}\
  \bibnamefont {Potvliege}},\ }\bibfield  {title} {\enquote {\bibinfo {title}
  {Free-space propagation of ultrashort pulses: Space-time couplings in
  gaussian pulse beams},}\ }\href {https://doi.org/10.1080/09500349908231378}
  {\bibfield  {journal} {\bibinfo  {journal} {Journal of Modern Optics}\
  }\textbf {\bibinfo {volume} {46}},\ \bibinfo {pages} {1881--1891} (\bibinfo
  {year} {1999})}\BibitemShut {NoStop}%
\bibitem [{\citenamefont {Marceau}, \citenamefont {April},\ and\ \citenamefont
  {Pich\'{e}}(2012)}]{Marceau:12}%
  \BibitemOpen
  \bibfield  {author} {\bibinfo {author} {\bibfnamefont {V.}~\bibnamefont
  {Marceau}}, \bibinfo {author} {\bibfnamefont {A.}~\bibnamefont {April}},\
  and\ \bibinfo {author} {\bibfnamefont {M.}~\bibnamefont {Pich\'{e}}},\
  }\bibfield  {title} {\enquote {\bibinfo {title} {Electron acceleration driven
  by ultrashort and nonparaxial radially polarized laser pulses},}\ }\href
  {https://doi.org/10.1364/OL.37.002442} {\bibfield  {journal} {\bibinfo
  {journal} {Opt. Lett.}\ }\textbf {\bibinfo {volume} {37}},\ \bibinfo {pages}
  {2442--2444} (\bibinfo {year} {2012})}\BibitemShut {NoStop}%
\end{thebibliography}%

\end{document}